\documentclass[onecolumn,draftclsnofoot]{IEEEtran}

 \usepackage{cite}
\usepackage{amsmath,amssymb,amsfonts}
\usepackage{algorithmic}
\usepackage{graphicx}
\usepackage{subcaption}
\usepackage{textcomp}
\usepackage{xcolor}
\usepackage{mathtools} 
\usepackage{booktabs} 

\usepackage{tikz} 
\usepackage{color}
\usepackage{siunitx}
\usepackage{hyperref}
\usepackage{amsfonts}
\usepackage{bbold}
\usepackage{tikz-cd}
\usetikzlibrary{shapes,snakes,shapes.geometric}
\usetikzlibrary{arrows,positioning}
\usepackage[linesnumbered,algoruled,boxed,lined,ruled]{algorithm2e}
\usetikzlibrary{fit,calc}
\usepackage{enumitem}
\newcommand*{\tikzmk}[1]{\tikz[remember picture,overlay,] \node (#1) {};\ignorespaces}
\newcommand{\boxit}[1]{\tikz[remember picture,overlay]{\node[yshift=3pt,fill=#1,opacity=.25,fit={(A)($(B)+(.975\linewidth,.8\baselineskip)$)}] {};}\ignorespaces}
\let\oldnl\nl
\newcommand{\nonl}{\renewcommand{\nl}{\let\nl\oldnl}}

\usepackage{amssymb,amsmath,amsthm,enumitem}

\newtheorem{assumption}{Assumption}

\newtheorem{definition}{Definition}
\newtheorem{example}{Example}
\newtheorem{proposition}{Proposition}

\usepackage{setspace}
 \newcommand{\citep}[1]{\cite{#1}}

\newcommand{\qeci}{{\sc \texttt{QECI}}}
\newcommand{\Tr}{\textbf{\textrm{Tr}}}

\begin{document}
\title{Quantum Entropic Causal Inference}

\author{\IEEEauthorblockN{Mohammad Ali Javidian, \and Vaneet Aggarwal, \and Fanglin Bao, \and  Zubin Jacob }\\
\IEEEauthorblockA{\textit{School of Electrical and Computer Engineering} \\
\textit{Purdue University}\\
\{mjavidia, vaneet, baof, zjacob\}@purdue.edu}
}

\maketitle
\begin{abstract}

The class of problems in causal inference which seeks to isolate causal correlations solely from observational data even without interventions has come to the forefront of machine learning, neuroscience and social sciences. As new large scale quantum systems go online, it opens interesting questions of whether a quantum framework exists on isolating causal correlations without any interventions on a quantum system. We put forth a  theoretical framework for merging quantum information science and causal inference by exploiting entropic principles. At the root of our approach is the proposition that the true causal direction minimizes the entropy of exogenous variables in a non-local hidden variable theory. The proposed framework uses a quantum causal structural equation model to build the connection between two  fields: entropic causal inference and the quantum marginal problem. First, inspired by the definition of geometric quantum discord, we fill the gap between classical and quantum conditional density matrices to define quantum causal models.  Subsequently, using a greedy approach, we develop a scalable algorithm for quantum entropic causal inference unifying classical and quantum causality in a principled way.  We apply our proposed algorithm to an experimentally relevant scenario of identifying the subsystem impacted by noise starting from an entangled state. This successful inference on a synthetic quantum dataset can have practical applications in identifying originators of malicious activity on future multi-node quantum networks as well as quantum error correction. As quantum datasets and systems grow in complexity, our framework can play a foundational role in bringing observational causal inference from the classical to the quantum domain.


\end{abstract}

\section{Introduction}\label{sec:intro}
The state-of-the-art method for causal inference is randomized experiments. Randomized controlled trials (RCT) provide a rigorous framework to examine cause-effect relationships between an intervention and outcome in clinical research \citep{hariton2018randomised} and many of the social sciences \citep{morgan2015counterfactuals}. In many cases, however, such a direct approach is
not possible due to expense, infeasibility, or ethical considerations. Instead, investigators have
to rely on observational studies alone to infer causality. However, causal inference solely from observational data is an ambitious and difficult task. 

One of the fundamental questions in causal analysis is to identify causal direction when cause and effect can be inferred from statistical information,
encoded as a joint probability distribution, obtained under normal, intervention-free measurement. Recent advances in computer science have made automated reasoning about cause and effect possible \citep{hoyer2008nonlinear,stegle2010probabilistic,cai2018causal,peters2011causal,peters2016causal,etesami2016interventional,mooij2016distinguishing,janzing2010causal,lee2017causal}.  This is a disruptive change in scientific methodology, but challenges and open problems exist \citep{pearl2018book}. 
The state-of-the-art technique for inferring the causal direction between two ordinal (or categorical) variables from observational data can be found in \citep{peters2011causal,janzing2012information,kaltenpoth2019we,Murat2017}. 

The approach in \citep{peters2011causal} is based on additive noise models (ANMs). In an ANM, the effect variable $Y$ is a linear function ($f$) of the cause variable $X$ plus an additive noise $E$ that is
independent of $X$, i.e., $Y=f(X)+E, E \perp\!\!\!\perp X$. This method suggests that if there is an ANM from $X$ to $Y$, but not vice versa, then $X\to Y$ is the causal direction. The method in \citep{janzing2012information,kaltenpoth2019we} is an information-theoretic approach based on Kolmogorov complexity. In short, this approach suggests that $X \to Y$ is only acceptable as causal direction if the shortest description of
$P_{ X,Y}$ is given by separate descriptions of $P_X$ and $P_{Y|X}.$ In other words,  first encoding of the true cause (i.e., $X$), and then the effect (i.e., $Y$) given that cause, results in a shorter
description\footnote{Here description length is understood in the sense of algorithmic information (“Kolmogorov complexity”) \citep{kolmogorov1965three}.} than other encoding of the observed variables.  The work that is most similar to ours in spirit is \citep{Murat2017}, which drops the restrictive additive noise assumption. This method applicable for classical systems, called entropic causal inference, is based on two main assumptions: (1) Every exogenous variable is a direct parent of at most one variable in the model (endogenous variable). This assumption is called causal sufficiency \citep{Pearl09}. (2) The entropy of the exogenous variable is small in the true causal direction.
This assumption (conjecture) was empirically validated in \citep{Murat2017}. We give a brief review of entropic causal inference in the classical context for the broad reader  in Appendix \ref{sec:classic}.


The problem of determining causal relations in quantum systems has been gaining attention \citep{gill2014statistics,pienaar2015graph,allen2017quantumb}, and new algorithms for quantum causal discovery have been designed \citep{fitzsimons2015quantum,ried2015quantum,giarmatzi2019quantum,chiribella2019quantum}. However, none of these methods can be used without  interventions from an experimentalist. In other words, they infer a causal model based on both observational and interventional data (i.e., the outcome of some experiments). Although causal inference can be improved
by considering interventional data, interventions are constrained by the thermodynamics of measurement and feedback in open
systems \citep{gachechiladze2020quantifying}. More importantly, as shown in \citep{milburn2018classical}, the perfect interventions characterised by
Pearl's do-calculus \citep{Pearl09} are physically impossible in
quantum systems \citep{milburn2018classical}. Since interventions are imprecise, costly, and in some cases impossible, inferring causal relationships from observational data alone is an important but challenging task in quantum causality, and \citep{Chaves2014UAI,chaves2014causal,chaves2015information} made early advances for this problem using information theoretical generalization of Bell’s inequalities and causal directed acyclic graphs (DAGs) in the quantum domain. However, this  approach does not differentiate between statistically equivalent DAGs, and in particular, cannot determine causal direction between two quantum systems since the two DAGs $A \to B$ and $A\gets B$ are statistically equivalent where $A$ and $B$ are two quantum systems. In this paper, we address the fundamental causal inference problem that involves only two quantum systems $A$ and $B$. Our goal is to address the frontier problem of quantum causal inference without any interventional data.

In this paper, we introduce a theoretical framework to merge quantum information science with causal inference using entropic principles. We first exploit a geometric measure of quantum discord to define an instance conditional density matrix. This physical quantity is used to construct quantum causal structural equations in a non-local hidden variable theory. The main proposition we exploit is that the true causal direction can be determined by finding the one which minimizes the entropy of exogenous variables. To find this exogenous variable with minimum entropy, we build the connection to the quantum marginal problem.  Although this problem is NP-hard in general, we adopt heuristic classical algorithms in the literature to the quantum domain for solving this problem. We formulate and numerically solve the minimum-entropy quantum marginal problem as an optimization problem, and we propose a scalable greedy minimization algorithm, as a counterpart of its classical version, to solve this problem.  Our method unifies classical and quantum causal inference in a principled way.




We put forth an experimental scheme that can be used to verify our theoretical framework. We consider a minimalistic model of an unknown message (possibly encrypted) with unknown origin in a two-node quantum network, where nodes are a coexisting set of quantum systems for which a joint density matrix can be defined \citep{chaves2015information,weilenmann2017analysing, weilenmann2020analysing}. Entangled quantum subsystems are used, where  one of the subsystems are communicated over a noisy channel (e.g., optical fiber)  to create such coexisting set of quantum systems.   We prove that only using the joint density matrix of two quantum systems, we can identify the originator of the message (i.e., the sub-system that did not encounter the noisy channel). To verify the validation of \qeci, we use realistic quantum noisy links such as quantum symmetric channel and depolarizing channel (valid for quantum networking and quantum communications). Moreover, we show that entropic causal inference technique cannot be mapped directly from quantum to classical framework in general, and it may result in erroneous outcomes. 

Our work deals with quantum generalizations of causal structures considering the absence of latent common causes\footnote{The absence of latent common causes (latent confounders) is called the causal sufficiency assumption \citep{sgs}.}. These structures can be shown as a directed acyclic graph (DAG), where nodes are quantum systems, and edges are quantum operations\footnote{In the context of quantum computation \citep{hogg1996quantum}, a quantum operation is called a quantum channel.}. However, the key theoretical distinction between an entirely classical causal structure and a quantum casual structure is the concept of coexistence. Because of the impossibility of cloning, the outcomes and the quantum systems that led to them do not exist simultaneously. If a system $A$ is measured to produce $B$, then $\rho_{AB}$ is not defined and hence neither is the entropy of $\rho_{AB}$ \citep{weilenmann2017analysing}.  

\section{Instance Conditional Density Matrix}\label{sec:conditionaldensity}
In this section, we define the concept of instance conditional density matrices as the extension of the conditional density matrices for a given joint density matrix $\rho_{XY}$, which is crucial to build our framework of entropic causal inference, as discussed in the next sections. For this purpose, we first review the formalism of quantum conditional states as a generalization of classical probability theory. Then, we define the notion of instance conditional density matrix, inspired by the another quantum information metric, i.e., quantum discord. Finally, we show that our definition of instance conditional density matrices is consistent with the EPR paradox.   

Quantum theory can be understood as a non-commutative generalization of classical probability theory wherein probability measures are replaced by density
operators \citep{leifer2013towards}. Analogies between the classical theory of Bayesian inference and the conditional states formalism for quantum theory are listed in Table \ref{table:analogies}.
\begin{table*}[!ht]
\caption{Analogies between classical and quantum formalism}
\centering 
\begin{tabular}{|l c l|} 
\hline 
\textbf{Classical Probability} &  & \textbf{Quantum Theory}\\ 
\hline \hline
probability distribution $p(X)$ & & density operator (matrix) $\rho_A$\\
\hline 
joint distribution $p(X,Y)$& & joint density $\rho_{AB}$\\
\hline
marginal distribution $p(X)=\sum_Yp(X,Y)$&& partial trace $\rho_A=\Tr_B(\rho_{AB})$\\
\hline
conditional probability $p(X|Y)=p(X,Y)/p(Y)$&& conditional density $\rho_{A|B}=(\rho_B^{-1}\otimes I)*\rho_{BA}$\\
\hline
instance conditional probability && \textbf{instance  conditional density matrix} \\
$p(X|Y=y)=\frac{p(X,Y=y)}{\sum_x p(X,Y=y)}$&& $\rho_{A|B=|b\rangle}=\frac{\Tr_B\{\rho_{BA}\star |b\rangle\langle b|\}}{trace\{\Tr_B\{\rho_{BA}\star |b\rangle\langle b|\}\}}$\\
\hline
\end{tabular}
\label{table:analogies} 
\end{table*}

Quantum conditional densities are a generalization of classical conditional probability distributions. However, to generalize conditional probabilities to the quantum case, several  approaches have been proposed in the literature. The three following generalizations are the best known in the literature of quantum information: (1) quantum conditional expectation \citep{umegaki1962conditional}, (2) quantum conditional amplitude operator \citep{cerf1997negative,cerf1999quantum}, and (3) quantum conditional states \citep{leifer2007conditional,leifer2013towards}.  Arguably, quantum conditional states are the most useful generalization
of conditional probability from the point of view of practical applications.

\begin{figure}[ht] 
  \begin{subfigure}[b]{0.475\linewidth}
    \centering
    \includegraphics[width=0.75\linewidth]{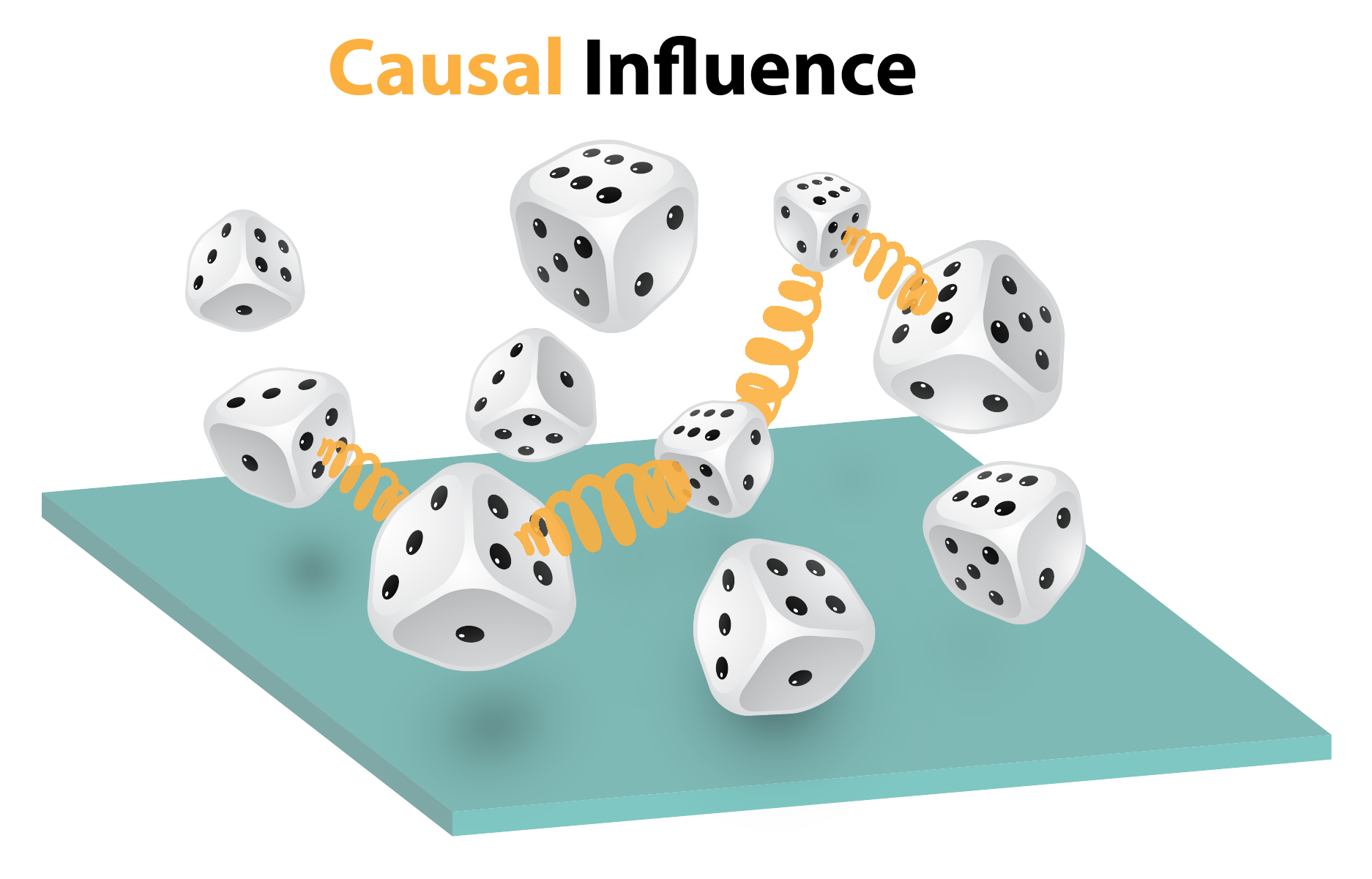} 
    \caption{We put forth a theory of causal inference for quantum systems using only observational data. For the broad audience, we present here a simple schematic to capture the main idea. If one has access only to observations of the final result from N dice, it is a frontier problem to analyze if a few elements influenced the other's outcomes. In this case, analyzing the entropy of the system provides insight even without time series data.} 
    \label{fig7:a} 
    \vspace{4ex}
  \end{subfigure}
  \hfill
  \begin{subfigure}[b]{0.475\linewidth}
    \centering
    \includegraphics[width=\linewidth]{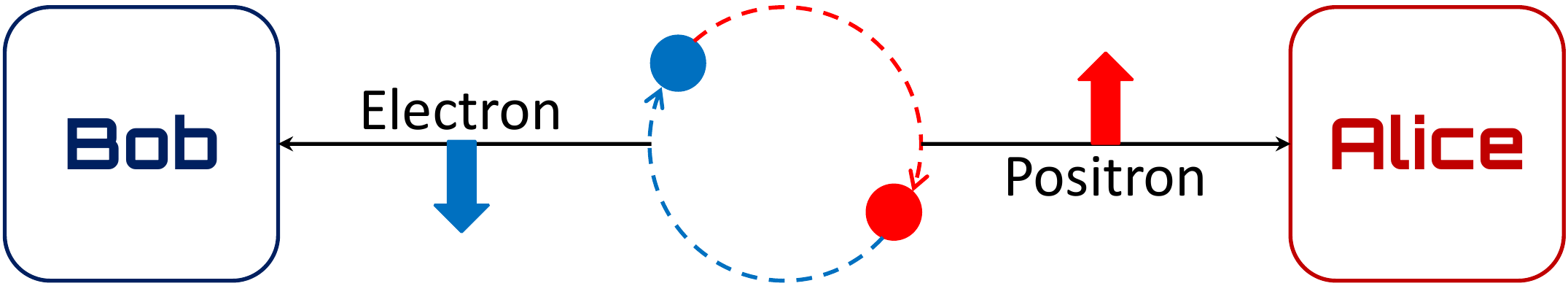}
    \caption{We build quantum causal structural equations using the instance conditional density matrix (ICDM). We show this ICDM captures the result of an electron-positron system in a spin-singlet state consists of two spin $\frac{1}{2}$ particles moving in opposite directions.}\label{fig:positronium} 
    \vspace{4ex}
  \end{subfigure}
  \caption{Building models of quantum entropic causal inference using instance conditional density matrices}
  \label{fig1}
\end{figure}

Despite the elegance of the idea and decades of efforts, the analogy between quantum conditional densities and classical conditional probability as it has been developed so far is not complete.
For example, consistent with the definition of conditional probability of events when $X$ is the
event $A = a$ and $Y$ is the event $B = b$, the conditional probability distribution
of $A$ given $B=b$ is defined as $$p(A|B=b)=\frac{p(A,B=b)}{\sum_x p(A,B=b)}$$

We will see that the quantum counterpart of this concept plays a significant role in the design of our entropic quantum causal inference algorithm. However, none of well-known generalizations of conditional probabilities in the literature of quantum computing has defined this notion where $B$ is in a given quantum state. To address this point, we first briefly introduce the concept of quantum discord and its connection to the concept of conditional density matrices. In quantum information theory, quantum discord is a measure of nonclassical correlations between two subsystems of a quantum system. However, the quantum discord between two quantum systems is not trivial to calculate, and except for some special classes of states there is no closed form solution for quantum discord. To overcome this difficulty, Daki\'{c} \cite{dakic2010necessary} introduced geometric quantum discord that is based on the Hilbert-Schmidt distance between the density matrix $\rho_{AB}$ and its closest classical state. For this purpose, conditional density matrices for single qubits have been used to estimate geometric quantum discord in the literature. 
For a given joint density matrix $\rho_{AB}$, the reduced density matrix $\rho_A$ given the measurement $B=|b\rangle^{obs}$ of the quantum system $B$, denoting by $\rho_{A|B=|b\rangle^{obs}}$. This quantity was used in \citep{dakic2010necessary} for defining quantum mutual information as follows:
$$Q_A(\rho_{AB})=S(\rho_B)-\min_{\{E_k\}}\sum_k p_kS(\rho_{B|k}),$$
where $S(\rho)=-\textrm{trace}(\rho\log \rho)$ is the von Neumann entropy and $\rho_{B|k}=\Tr_A(E_k\otimes\mathbb{1}_B\rho_{AB})/\Tr(E_k\otimes\mathbb{1}_B\rho_{AB})$ is the state of $B$ conditioned on outcome $k$ in $A$, and $\{E_k\}$ represents the set
of positive operator valued measure elements \citep{dakic2010necessary}. Daki\'{c} defines the quantum discord as the discrepancy between the two measures of information as follows:
\begin{equation}\label{eq:discord}
    D_A(\rho_{AB}) = I_Q(\rho_{AB}) -Q_A(\rho_{AB}),
\end{equation}
where $I_Q(\rho_{AB})=S(\rho_A)+S(\rho_B)-S(\rho_{AB})$.
Inspired by Daki\'{c}'s work, we formally define the instance conditional density matrices for an arbitrary quantum system $B$ as follows.

\begin{definition}[Instance Conditional Density Matrix]\label{def:cond_density}
For a given joint density matrix $\rho_{AB}$, the reduced density matrix $\rho_A$ given the specific measurement instance post-observation $B=|b\rangle^{obs}$ of the quantum system $B$, i.e., $\rho_{A|B=|b\rangle^{obs}}$, is given  as
\begin{equation}
\rho_{A|B=|b\rangle^{obs}}=\frac{\Tr_B\{\rho_{BA}\star |b\rangle^{obs\enskip obs}\langle b|\}}{trace\{\Tr_B\{\rho_{BA}\star |b\rangle^{obs\enskip  obs}\langle b|\}\}}
\end{equation}
where the $\star$-product is defined by $M\star N:=(N^{1/2}\otimes I)M(N^{1/2}\otimes I)$, and the trace of a square matrix $A$, denoted $trace(A)$, is defined to be the sum of elements on the main diagonal of $A$ (for simplicity we drop the superscript $^{obs}$  from now on). Thus, $\rho_{BA}\star |b\rangle\langle b|=((|b\rangle\langle b|)^{1/2}\otimes I)*\rho_{BA}* ((|b\rangle\langle b|)^{1/2}\otimes I)$.
\end{definition}

The instance conditional density matrix $\rho_{A|B=|b\rangle}$, defined in Definition \ref{def:cond_density}, is always a valid density matrix. Note,
$\rho_{BA}\star |b\rangle\langle b|$ is a semidefinite positive operator because it is of the form $VV^\dagger$, where $V=((|b\rangle\langle b|)^{1/2}\otimes I)\rho_{BA}^{1/2}$. Also, note that  the partial trace of a positive semidefinite matrix is a positive semidefinite matrix \citep{filipiak2018properties}. However, the numerator of $\rho_{A|B=|b\rangle}$ is not a density operator because it does not have trace one. Normalization guarantees that $\rho_{A|B=|b\rangle}$ is a valid density matrix.


In the following, we show that our definition of instance conditional density matrix is consistent with the EPR paradox. In order to see that, consider an electron-positron system in a spin-singlet state (i.e., a system of two spin $\frac{1}{2}$ particles moving in opposite directions, as shown in Figure \ref{fig:positronium}), in which the total spin of the system is equal to zero.

The operators corresponding to the spin along the $x, y$, and $z$ direction, denoted $S_x, S_y$, and $S_z$ respectively, can be represented using the Pauli matrices \citep{Sakurai2021}:
\begin{equation}\label{eq:spinoperator}
    S_x=\frac{\hbar}{2}\begin{bmatrix}
			0 & 1 \\
			1 & 0
		\end{bmatrix},S_y=\frac{\hbar}{2}\begin{bmatrix}
			0 & -i \\
			i & 0
		\end{bmatrix}, S_z=\frac{\hbar}{2}\begin{bmatrix}
			1 & 0 \\
			0 & -1
		\end{bmatrix} 
\end{equation}
where $\hbar$ is the Planck constant divided by $2\pi$. The eigenstates of $S_x$, i.e., $|x^+\rangle$ and $|x^-\rangle$ are represented  as  normalized eigenvectors $\frac{1}{\sqrt{2}}\begin{bmatrix}
			1 & 1
		\end{bmatrix}^T$ and $\frac{1}{\sqrt{2}}\begin{bmatrix}
			1 &
			-1
		\end{bmatrix}^T$, respectively. The eigenstates of $S_z$, i.e., $|z^+\rangle$ and $|z^-\rangle$ are represented  as  normalized eigenvectors $\begin{bmatrix}
			1 &
			0
		\end{bmatrix}^T$ and $\begin{bmatrix}
			0 &
			1
		\end{bmatrix}^T$, respectively. The eigenstates of $S_y$, i.e., $|y^+\rangle$ and $|y^-\rangle$ are represented  as  normalized eigenvectors $\frac{1}{\sqrt{2}}\begin{bmatrix}
			1 &
			i
		\end{bmatrix}^T$ and $\frac{1}{\sqrt{2}}\begin{bmatrix}
			1 &
			-i
		\end{bmatrix}^T$, respectively. The state ket (i.e., the spin wave function) can be written as:
		\begin{equation}\label{eq:stateketz}
		    |\psi\rangle=\frac{1}{\sqrt{2}}(|z_e^+\rangle|z_p^-\rangle-|z_e^-\rangle|z_p^+\rangle)
		\end{equation}
where $|z_e^+\rangle|z_p^-\rangle$ means that electron is in the spin-up state and positron is in the spin-down state. If we choose the $x$-direction as the axis of quantization, we can rewrite spin-singlet ket as:
\begin{equation}\label{eq:stateketx}
		    |\psi\rangle=-\frac{1}{\sqrt{2}}(|x_e^+\rangle|x_p^-\rangle-|x_e^-\rangle|x_p^+\rangle)
\end{equation}
Apart from the overall sign, Eq. \eqref{eq:stateketz} and \eqref{eq:stateketx} indicate that spin-singlet states have no preferred direction in space. Also, if we choose the $y$-direction as the axis of quantization, we can rewrite spin-singlet ket as:
\begin{equation}\label{eq:statekety}
		    |\psi\rangle=\frac{i}{\sqrt{2}}(|y_e^+\rangle|y_p^-\rangle-|y_e^-\rangle|y_p^+\rangle)
\end{equation}

The reason for having equations \eqref{eq:stateketx}, and \eqref{eq:statekety} is that for a single spin $\frac{1}{2}$ system the $S_x$ eigenkets, $S_y$ eigenkets, and $S_z$ eigenkets are related as follows \citep{mohrhoff2011world}:
\begin{equation}\label{eq:eigenketseqiv}
    |x^\pm\rangle=\frac{1}{\sqrt{2}}(|z^+\rangle\pm|z^-\rangle),\quad |y^\pm\rangle=\frac{1}{\sqrt{2}}(|z^+\rangle\pm i|z^-\rangle)
\end{equation}
So, the density matrix $\rho_{e,p}$ for the spin singlet state $|\psi\rangle$ is:
\begin{equation}\label{eq:densitysinglet}
    \rho_{e,p}=|\psi\rangle\langle\psi|=\begin{bmatrix}
			0&0&0&0 \\
			0&1&-1&0 \\
			0&-1&1&0 \\
			0&0&0&0
		\end{bmatrix}
\end{equation}
It is not difficult to verify that $\rho_p=\rho_e=\frac{I}{2}$, which means both $\rho_p$ and $\rho_e$ are maximally mixed states. Consider Figure \ref{fig:positronium}, and assume that Alice specializes in measuring $S_z$ of positron, while Bob specializes in measuring $S_z$ of electron. Let us assume that Alice finds $S_z$ to be positive (state-up) for positron. Then Alice can predict, even before Bob performs any measurement, the outcome of Bob’s measurement with certainty: Bob must find $S_z$ to be negative (state-down) for electron. Here, we show that using our definition of the instance conditional density matrices (Definition \ref{def:cond_density}) gives the same result. Note that since $\rho_{e,p}$ is symmetric,  $\rho_{e,p}=\rho_{p,e}$. Using Definition \ref{def:cond_density}, we have:
\begin{equation}\label{eq:cond_density_positronium}
    \rho_{e|p=|z_p^+\rangle}=\frac{\Tr_p(((|z_p^+\rangle\langle z_p^+|)^{1/2}\otimes I_2)\rho_{p,e}(((|z_p^+\rangle\langle z_p^+|)^{1/2}\otimes I_2)))}{trace(\Tr_p(((|z_p^+\rangle\langle z_p^+|)^{1/2}\otimes I_2)\rho_{p,e}(((|z_p^+\rangle\langle z_p^+|)^{1/2}\otimes I_2))))}=\begin{bmatrix}
    0 & 0\\
    0 & 1
    \end{bmatrix}=|z_e^-\rangle\langle z_e^-|
\end{equation}

Eq. \eqref{eq:cond_density_positronium} says: for the given joint density matrix $\rho_{e,p}$, the reduced density matrix $\rho_e$ given the specific measurement instance post-observation $|z_p^+\rangle$ of the quantum subsystem $p$, i.e., $\rho_{e|p=|z_p^+\rangle}$, is $|z_e^-\rangle$. In other words, if Alice makes a measurement and finds that $S_z$ is positive (state-up) for positron, then Bob must find $S_z$ to be negative (state-down) for electron. From Eq. \eqref{eq:eigenketseqiv}, we obtain:
\begin{equation}\label{eq:eqivalent}
    |z_e^-\rangle=\frac{1}{\sqrt{2}}(|x_e^+\rangle+|x_e^-\rangle), \quad |z_e^-\rangle=\frac{-i}{\sqrt{2}}(|y_e^+\rangle-|y_e^-\rangle)
\end{equation}
which means if Alice measures $S_z$ and Bob measures $S_x$ (or $S_y$), there is a completely random correlation between
the two measurements, and Bob has 50\% chance of getting $|x_e+\rangle$ or $|x_e^-\rangle$ ($|y_e^+\rangle$ or $|y_e^-\rangle$). Equations \eqref{eq:cond_density_positronium} and \eqref{eq:eqivalent} together confirm that our definition of instance conditional density matrix (Definition \ref{def:cond_density}) is consistent with the EPR paradox.
Notice that Alice and Bob can be miles apart with no possibility of communications or mutual interactions.

\section{Non-local hidden variable theory for quantum causal structural equations}\label{intro:str}

Consider two  quantum subsystems $A$ and $B$.
Assume that density matrices $\rho_A$ and $\rho_B$ are defined on Hilbert spaces $\mathcal{H}_A$ and $\mathcal{H}_B$,  respectively. Following the key assumptions used for determining the causal direction in  \citep{Murat2017} (see section \ref{sec:classic} for a brief review), we define the causal graph direction $A\to B$ based on the following property that expresses non-local hidden variable theory:
\begin{equation}\label{quant_st1}
    \rho_{B|A=|a\rangle^{obs}} = f(|a\rangle^{obs},\rho_{E})
\end{equation}
Here, the instance conditional density matrix $\rho_{B|A=|a\rangle^{obs}}$ is related for all $|a\rangle^{obs}$ to an unknown exogenous density matrix $\rho_{E}$ which does not depend on $|a\rangle$ (for simplicity we drop $^{obs}$  from now on). We choose the exogenous density matrix $\rho_{E}$ as the one which minimizes the entropy of $\rho_E$ such that \eqref{quant_st1} is satisfied. To determine the cause-effect direction, we exploit the proposition that the entropy of the exogenous density matrix is minimized in the true causal direction. This framework in terms of density matrices generalizes the random variable framework of defining the causality direction in   \citep{Murat2017}. We will discuss this framework in detail in section \ref{sec:formalQECI}.

We consider an illustrative example to show that the quantum causal structural equation with non-local hidden variables is indeed satisfied for a noisy channel. We have two entangled quantum bits and one is sent over a quantum channel. The system $\rho_{AB}$ is given as
\begin{equation}
    \rho_{AB} = \frac{1}{2}(1-p)(|00\rangle\langle00| + |11\rangle\langle11|) + \frac{1}{2}p(|01\rangle\langle01| + |10\rangle\langle10|) \label{eq:introeg}
\end{equation}
Here the system $(A,B)$ is prepared in a superposition of density matrices $|00\rangle\langle00|$ and $|11\rangle\langle11|$. Then, the subsystem $B$ is passed through a Pauli bit-flip error channel with probability $p$. In this case, we expect the causal direction as $A\to B$. We now provide the structure of the form of Eq. \eqref{quant_st1}. For $A=|a\rangle$ with $a\in \{0,1\}$, we have 
\begin{equation}
    \rho_{B|A=|a\rangle} = (1-p)|a\rangle\langle a|  + p(\sigma_X|a\rangle)(\sigma_X|a\rangle)^\dagger, \label{instance_intro}
\end{equation}
where $\sigma_X=\begin{bmatrix}
			0 & 1 \\
			1 & 0\\
		\end{bmatrix}.$ 
Let $M =|a\rangle^\dagger \otimes I_2,$ where $I_2$ is an identity matrix of size $2$. 		For the specific noisy channel, we have the form of the exogenous density matrix to be 
\begin{equation}
\rho_E = \frac{1}{2}diag([1-p, p, p, 1-p]),
\end{equation}
we obtain the causal structural function 
\begin{equation}
f(|a\rangle,\rho_{E}) = 2M\rho_E M^\dagger = \rho_{B|A=|a\rangle} ,
\end{equation}
Thus, the above noisy transmission of $B$ satisfies the quantum causal structural equation with the noise pattern for  the system. We also note that in this example, the two subsystems $A$ and $B$ co-exist to be able to define the joint and instance-conditional density matrices. We further note that based on the joint density matrix in \eqref{eq:introeg} and instance conditional density matrix in \eqref{instance_intro}, measurement of one subsystem impacts the other, even when at a distance. Thus, this is consistent with the quantum literature where the distant events are not independent thus violating the local hidden-variable theory. A more general investigation of quantum causal structural equations and non-local hidden variable theory will be provided in future work.

\section{Framework of Quantum Entropic Causal Inference}\label{sec:formalQECI}
As  in Section \ref{intro:str}, we consider two  quantum subsystems $A$ and $B$. Assume that density matrices $\rho_A$ and $\rho_B$ are defined on Hilbert spaces $\mathcal{H}_A$ and $\mathcal{H}_B$,  respectively. We define the causal graph direction $A\to B$ based on the following property:
\begin{equation}\label{quant_st}
    \rho_{B|A=|a\rangle} = f(|a\rangle,\rho_{E})
\end{equation}
for all $|a\rangle$, some exogenous density matrix $\rho_{E}$ which does not depend on $|a\rangle$, and the instance conditional density matrix $\rho_{B|A=|a\rangle}$. The exogenous  density matrix $\rho_{E}$  is the one that minimizes the joint entropy of $\rho_{B|A=|a\rangle}$ for all $|a\rangle$'s. 
Formally, this means that 
\begin{equation}
S(\rho_E)\approx\min S(\rho_{B|A=|a_1\rangle,\cdots,{B|A=|a_n\rangle}}),
\end{equation}
for $|a_1\rangle$, $\cdots$,$|a_n\rangle$ as the possible measurement outcomes of $A$. Even though the individual densities $\rho_{B|A=|a_i\rangle}$ are known, the minimum is over all possible couplings of these density matrices. Since $E$ is in the direction of $A\to B$, we label $S(\rho_A)+S(\rho_E)$ as $S(A\to B)$. This determines the entropy of the exogenous density matrix in the direction from $A\to B$.



We now describe the key proposition of quantum causal inference. Given a set of quantum causal structural equations that describe the system, the true causal direction can be ascertained as follows: 
\begin{proposition}\label{qmainassump}
Entropy of the exogenous  density matrix $\rho_{E}$ is small
in the true causal direction. In other words, the exogenous density matrix has
lower entropy in the true causal direction than the entropy in the wrong direction. So, if $A\to B$ is the right causal direction, we formally have: $S(A\to B)<S(B\to A)$.
\end{proposition}

Using this proposition, we can compute the entropy of the exogenous density matrices in both directions, and find the smaller to determine the causal direction. The detailed algorithm for achieving is given in the following section.


\section{Minimum Entropy Quantum Marginal}\label{sec:qcoupling}
We first relate the problem of finding the density matrix $\rho_E$ with minimum entropy in the equation (\ref{quant_st1}) to the problem of minimum-entropy quantum couplings (also known as quantum marginal problem) as an optimization problem. This is one of the widely studied problems in quantum information theory for characterizing the set of possible density matrices given the marginals \citep{zeng2019quantum} which is also known as  the $N$-representability problem in quantum chemistry \citep{klyachko2006quantum,liu2007quantum}.  The general problem is of interest in many-body quantum simulation but is computationally intractable \citep{liu2006consistency}. In this section, we propose a greedy algorithm for solving this problem. 
In the next section, we show the fundamental role of this connection in solving the problem of causal inference via an entropic approach. 

\subsection{Finding $\rho_E$ with minimum entropy as the solution to minimum entropy quantum coupling}
Define the function $f_{|a_i\rangle}(\rho_E):= f(|a_i\rangle,\rho_E)$, for $i=1,\dots, n$. From Eq. (\ref{quant_st1}) we have: 
$\rho_{B|A=|a_i\rangle} = f(|a_i\rangle,\rho_{E})$. Also, define $\rho_{U_1,\dots,U_n}$ as a joint density matrix such that each $\rho_{U_i}=f_{|a_i\rangle}(\rho_E)$ can be obtained from $\rho_{U_1,\dots,U_n}$ via partial trace operators. Now, consider the following optimization problem: 
\begin{equation}\label{eq:findingRhoE}
    \begin{array}{cl}
     & \min_{\rho_{U_1,\dots,U_n}}S(\rho_{U_1,\dots,U_n})  \\
     & \textrm{subject to }\Tr_{U_1,\dots,U_{i-1},U_{i+1},\dots,U_n}(\rho_{U_1,\dots,U_n})=\rho_{U_i}, \forall i=1,\dots,n,
\end{array}
\end{equation}
where the von Neumann entropy of a quantum state $\rho$, defined as $S(\rho)=-\textrm{trace}(\rho\log \rho)$, is non-negative and concave. The formulated problem in (\ref{eq:findingRhoE}) is called Minimum Entropy Quantum Coupling Problem. 
Let $\rho_{U_i}=f_{|a_i\rangle}(\rho_E)$, and assume that $\rho_E$, $\rho_{U_1,\dots,U_n}$, and all $\rho_{U_i}$s are non-negative diagonal matrices for now. In the following, we show that $\rho_E$ with minimum entropy in Eq. (\ref{quant_st1}) is the solution to minimum entropy quantum coupling problem in (\ref{eq:findingRhoE}). 
By the definition of von Neumann entropy and the quantum conditional entropy we have:
\begin{align}\label{eq:classicalcoupling}
S(\rho_E)
&= \underbrace{S(\rho_E|\rho_{U_1,\dots,U_n})}_{\ge 0} +S(\rho_{U_1,\dots,U_n})-\underbrace{S(\rho_{U_1},\dots,\rho_{U_n}|\rho_E)}_{=0} \\
& \ge S(\rho_{U_1,\dots,U_n}).
\end{align}
So, we can find the best lower bound for the von Neumann entropy of the exogenous density matrix $\rho_E$ by solving the following optimization problem.
\begin{equation*}
    \begin{array}{cl}
     S(\rho_E)\ge& \min_{\rho_{U_1,\dots,U_n}}S(\rho_{U_1,\dots,U_n})  \\
     & \textrm{subject to }\Tr_{U_1,\dots,U_{i-1},U_{i+1},\dots,U_n}(\rho_{U_1,\dots,U_n})=\rho_{U_i}, \forall i=1,\dots,n,
\end{array}
\end{equation*}
It is not difficult to see that we are able to construct a density matrix $\rho_E$ that achieves this minimum. For this purpose, assume that the optimal joint density matrix is $\rho_{U_1,\dots,U_n}^*$, and without loss of generality assume that each $\rho_{U_i}$ has $m$ states. We can construct $\rho_{E}^*$ consists of $m^n$ states with states equal
to $\rho_{U_1,\dots,U_n}$ for each configuration of $\rho_{U_1,\dots,U_n}$. This $\rho_{E}^*$ has the same entropy as the joint entropy, since
state values are the same. So,
\begin{equation*}
    \begin{array}{cl}
     S(\rho_{E}^*)=& \min_{\rho_{U_1,\dots,U_n}}S(\rho_{U_1},\dots,\rho_{U_n})  \\
     & \textrm{subject to }\Tr_{U_1,\dots,U_{i-1},U_{i+1},\dots,U_n}(\rho_{U_1,\dots,U_n})=\rho_{U_i}, \forall i=1,\dots,n,
\end{array}
\end{equation*}
This means, in this case the problem of finding the exogenous density matrix $\rho_E$ with minimum von Neumann entropy given the joint density $\rho_{AB}$ is equivalent to the problem of finding the minimum entropy of $\rho_{U_1},\dots,\rho_{U_n}$, subject to the constraint that each $\rho_{U_i}$ is the instance density matrix $\rho_{B|A=|a_i\rangle}$ and partial trace of $\rho_{U_1},\dots,\rho_{U_n}$. In other words, $\rho_E$ is the solution of the minimum entropy quantum coupling.

We note that $\rho_{U_i}, \forall i=1,\dots,n$, can be diagonalized without impact results but $\rho_{U_1},\dots,\rho_{U_n}$ in general may not be diagonal and similarly $\rho_E$, which limits possibility of the following conditions $S(\rho_E|\rho_{U_1,\dots,U_n})\ge 0$ and $S(\rho_{U_1},\dots,\rho_{U_n}|\rho_E)=0$. Despite that, the optimal solution for the minimum entropy coupling problem in (\ref{eq:findingRhoE}) is an achievable point  for the problem of finding the density matrix $\rho_E$ with minimum entropy in Eq. (\ref{quant_st1}) that we use later to solve the problem of quantum entropic causal inference.     

The set of positive semidefinite matrices is a closed convex set in the space of symmetric matrices \citep{Lange2013}. Also, the solution set of a system of linear
equations is an affine space\footnote{An affine space is  like a vector space, except that no special choice of origin is assumed.}, and as a result a convex set \citep{boyd2004convex}. Thus, the constraints in Eq. (\ref{eq:findingRhoE}) define convex constraints for  minimum entropy coupling problem. This means that the minimum entropy quantum marginal problem is a concave minimization problem, subject to convex constraints. 

A large number of algorithms
have been developed for solving concave minimization problems. A survey and
comparison of methods can be found in \citep{heising1981survey,pardalos1986methods}. Details concerning solution approaches to solve concave minimization problems can be found in \citep{pardalos1987constrained,locatelli2013global,horst2013global,Tuy2016}. However, there are some important practical issues to use these approach for solving the minimum entropy quantum marginal: (1) scalability: This optimization problem is difficult to perform numerically because the boundary of the space of positive semidefinite matrices is hard to compute. (2) Most of optimization techniques do not guarantee global optimality, and it is quite possible to return a local optimum. In general, finding the global minimum for the minimum entropy quantum marginal is  difficult (NP hard): analytical methods are not applicable, and the use of a numerical solution strategies often does not lead to optimal solutions in polynomial time. To overcome these difficulties we propose an alternative approach in the following subsection.

\subsection{A Greedy Entropy Minimization Algorithm for Quantum Marginal Problem}\label{sec:greedyalg}
To succeed in dealing with the problems in solving the minimum-entropy quantum marginals, we propose a heuristic greedy algorithm that connects this problem to the classical minimum-entropy quantum couplings. In the next section, we show how to use this strategy for solving the problem of quantum causal inference. However, applicability of our proposed method goes beyond causal inference and can be used by the active community \citep{zeng2019quantum,schilling1507quantum,fritz2012entropic} that are working on quantum marginals. For simplicity, we consider $n=2$, while the approach directly works for general $n$.  
Given two density matrices $\rho_{U_1}$ and $\rho_{U_2}$ of two quantum systems ${U_1}$ and ${U_2}$ respectively, the minimum entropy quantum
coupling problem is to find the minimum-entropy joint density matrix $\rho_{{U_1}{U_2}}$ among all possible joint density matrices having $\rho_{U_1}$ and $\rho_{U_2}$ as partial traces. This problem is known to be NP-hard, even in the classical context \citep{kovavcevic2015entropy}. In this section, we propose  an algorithm that provides a feasible solution to the quantum marginal problem.


Any density operator defines a classical probability distribution: its eigenvalues are a probability distribution on the set of its eigenvectors \citep{rieffel2011quantum} ([Section 10.1.2]), formally we have:
\begin{eqnarray}
	\rho_{U_1}=\sum_ip_i|v_i\rangle\langle v_i|,\ p_i\ge 0,\ \Tr(\rho_{U_1})=\sum_ip_i=1,\label{eq:quantum2classic3}\\
	\rho_{U_2}=\sum_jq_j|w_j\rangle\langle w_j|,\ q_j\ge 0,\ \Tr(\rho_{U_2})=\sum_jq_j=1.\label{eq:quantum2classic4}
\end{eqnarray}

Here $|v_i\rangle$ denotes the normalized orthogonal eigenvectors of $\rho_{U_1}$, with $|v_i\rangle\langle v_i|$ the corresponding
orthogonal projector, and $p_i$ its eigenvalues, such that $\langle v_i|\rho_{U_1}|v_j\rangle = p_i\delta_{ij}$,  where $\delta_{ij}$ is the Kronecker symbol ($\delta_{ij} = 1$ for $i = j$ and 0 otherwise). The eigenvalue
$p_i$ represents the probability of finding the system in the pure quantum state $|v_i\rangle$. Similar statement holds for $|w_j\rangle$ and $\rho_{U_2}$. 


Our proposed algorithm (pseudo-code in Algorithm \ref{alg:qgreedy}) is composed of two steps.  In the first step, we compute the eigenvalue decomposition of the given density matrices $\rho_{U_1}$ and $\rho_{U_2}$. This helps give the eigenvectors $|v_i\rangle$, $|w_j\rangle$ as well as the probabilities $p_i$ and $q_j$. In the second step, we apply the greedy algorithm provided in \citep{kocaoglu2017entropic} on these probabilities $p_i$ and $q_j$ of the two random variables to determine the joint probabilities of the joint random variable. Let the joint probabilities returned by the algorithm be $r_{i,j}$ corresponding to the first variable having probability $p_i$ and the second variable having probability $q_j$. The joint entropy returned by the algorithm is defined as $\textrm{Joint\_Entropy}(r)=-\sum_{i,j}r_{i,j}\log(r_{i,j})$. We further note that the joint density matrix which has the entropy as above is
\begin{equation}
	\rho_E=\sum_{i,j}r_{i,j}|v_i\rangle|w_j\rangle(|v_i\rangle|w_j\rangle)^\dagger
\end{equation}
Since the above $\rho_E$ can be easily shown to be a valid density matrix, with $\textrm{Joint\_Entropy}(r)$, this $\rho_E$ is a feasible solution for the joint coupling problem. We further note that the classical coupling problem is a special case of the quantum marginal problem where the pure states $|v_i\rangle$, $|w_j\rangle$ are the possibilities of the classical variables. In this case, the quantum marginal problem results in the same joint entropy as in the classical coupling problem in \citep{kocaoglu2017entropic}. Thus, the proposed approach results in a feasible joint density matrix with the constraints in \eqref{eq:findingRhoE} satisfied.

We note that in step 2 of Algorithm \ref{alg:qgreedy}, there are alternative greedy algorithms  \citep{cicalese2017find,rossi2019greedy} that can be used in this step. These algorithms provide different guarantees, i.e., solutions that are local minimum \citep{Murat2017} and 1-bit approximation \citep{cicalese2017find}. In \citep{rossi2019greedy}, the authors proved that the
algorithm proposed in \citep{Murat2017} provides, in addition, a 1-bit approximation guarantee in the case of two variables. Moreover, Algorithm \ref{alg:qgreedy} can be extended to more than two quantum systems. In those cases, Joint Entropy Minimization Algorithm \citep{kocaoglu2017entropic} can be replaced by the one in \citep{Murat2017}. 



\section{\qeci: Proposed Algorithm for Quantum Entropic Causal Inference}\label{sec:qeci}
\begin{figure*}[!ht]
    \centering
    \includegraphics[width=\linewidth]{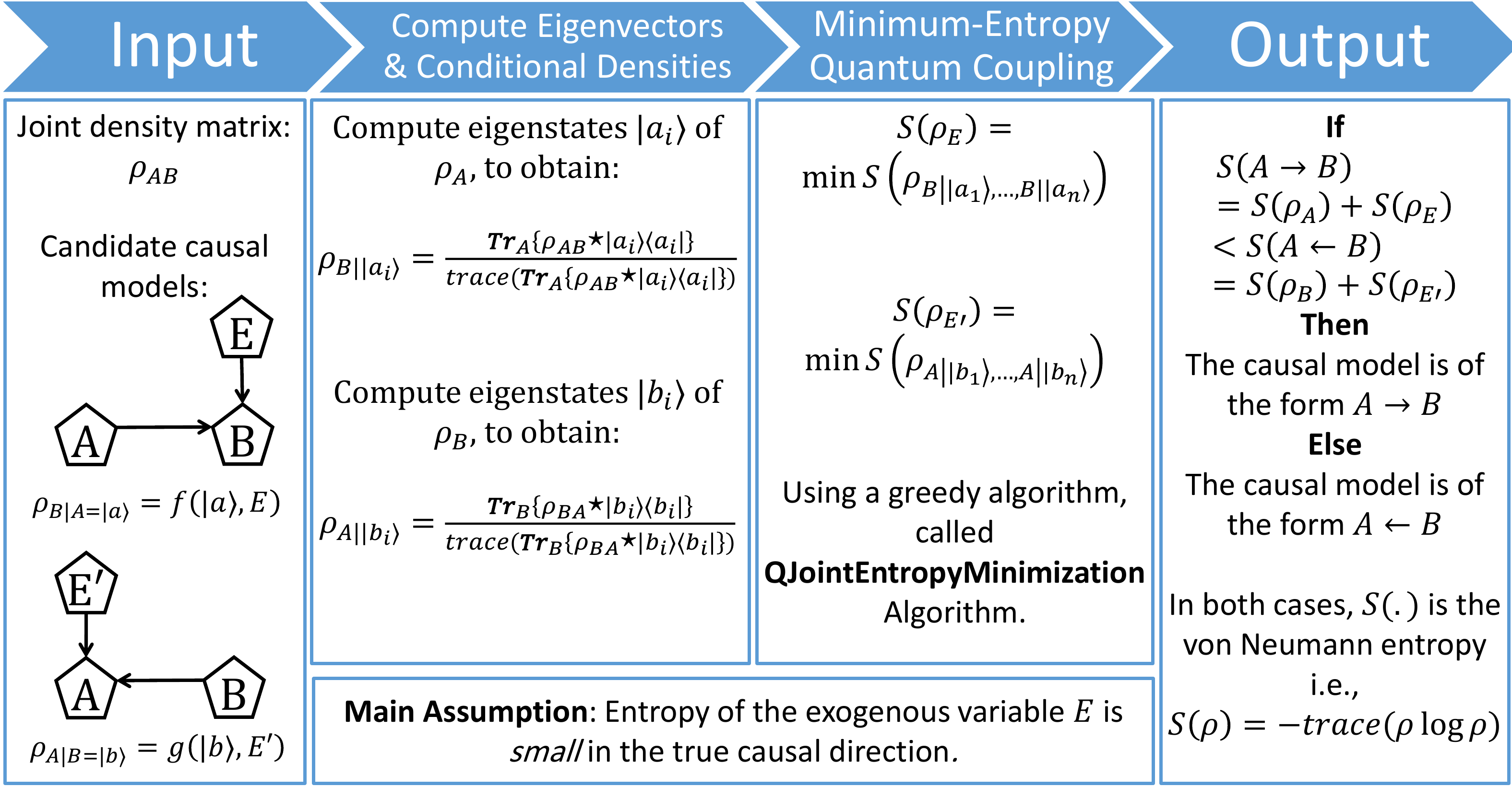}
    \caption{An overview of \qeci: an algorithm for \textbf{Q}uantum \textbf{E}ntropic \textbf{C}ausal \textbf{I}nference.}
    \label{fig:QECI}
\end{figure*}

In this section we give the algorithm of quantum entropic causal inference. This problem requires finding the minimum entropy of the exogenous density matrix which is equivalent to the minimum-entropy quantum coupling problem, as studied in Section \ref{sec:qcoupling}. Then, Assumption \ref{qmainassump} which states that the entropy of the exogenous density matrix is lower in the causal direction is used to determine the causal direction in the quantum entropic causal inference. The detailed algorithm is provided in this section using a density matrix approach. Further, an overview of our \textbf{Q}uantum \textbf{E}ntropic \textbf{C}ausal \textbf{I}nference (\qeci) algorithm is shown in Figure \ref{fig:QECI}, and Algorithm \ref{alg:qeci} summarizes the entire procedure of \qeci. 

\textbf{\qeci~Description.} For a given joint density matrix $\rho_{AB}$, \qeci~is able to discover the true causal direction between quantum subsystems $A$ and $B$ in four phases:\\
\textbf{Phase 1.} In this phase, we trace out $B$ and $A$ to obtain $\rho_A$ and $\rho_B$, respectively. Then, we compute $\rho_{BA}$ by reordering the entries of $\rho_{AB}$, appropriately.\\
\textbf{Phase 2.} Computing Eigenvectors and Conditional Densities: In this phase, we first find the eigenvectors of $\rho_A$ and $\rho_B$ to determine the eigenstates. These eigenstates are $|a_i\rangle$ and $|b_j\rangle$, respectively. These eigenstates provide a representation of the mixed states $\rho_A$ and $\rho_B$ into superposition of pure states.  Using these eigenstates,  Definition \ref{def:cond_density} is used to find the instance conditional density matrices.\\
\textbf{Phase 3.} Solve Quantum Marginal Problem for the Conditional Densities: In this step, we use the quantum marginal problem to determine the minimum-entropy quantum coupling between different $\rho_{B|A=|a_i\rangle}$s. Similarly, the quantum coupling between the different $\rho_{A|B=|b_j\rangle}$ is obtained.\\
\textbf{Phase 4.} Based on the minimum entropy coupling, we estimate minimum entropy $S(A\to B)$ and $S(A\gets B)$. Taking into account Assumption \ref{qmainassump}, if $S(A\to B)<S(A\gets B)$, \qeci~returns that the causal model is of the form $A\to B$, and $A\gets B$ otherwise.

\section{Evaluation on Quantum Cause-Effect Synthetic Data}\label{sec:examples}
\begin{figure}[ht] 
  \begin{subfigure}[t]{0.47\textwidth}
    \centering
    \includegraphics[width=.8\linewidth]{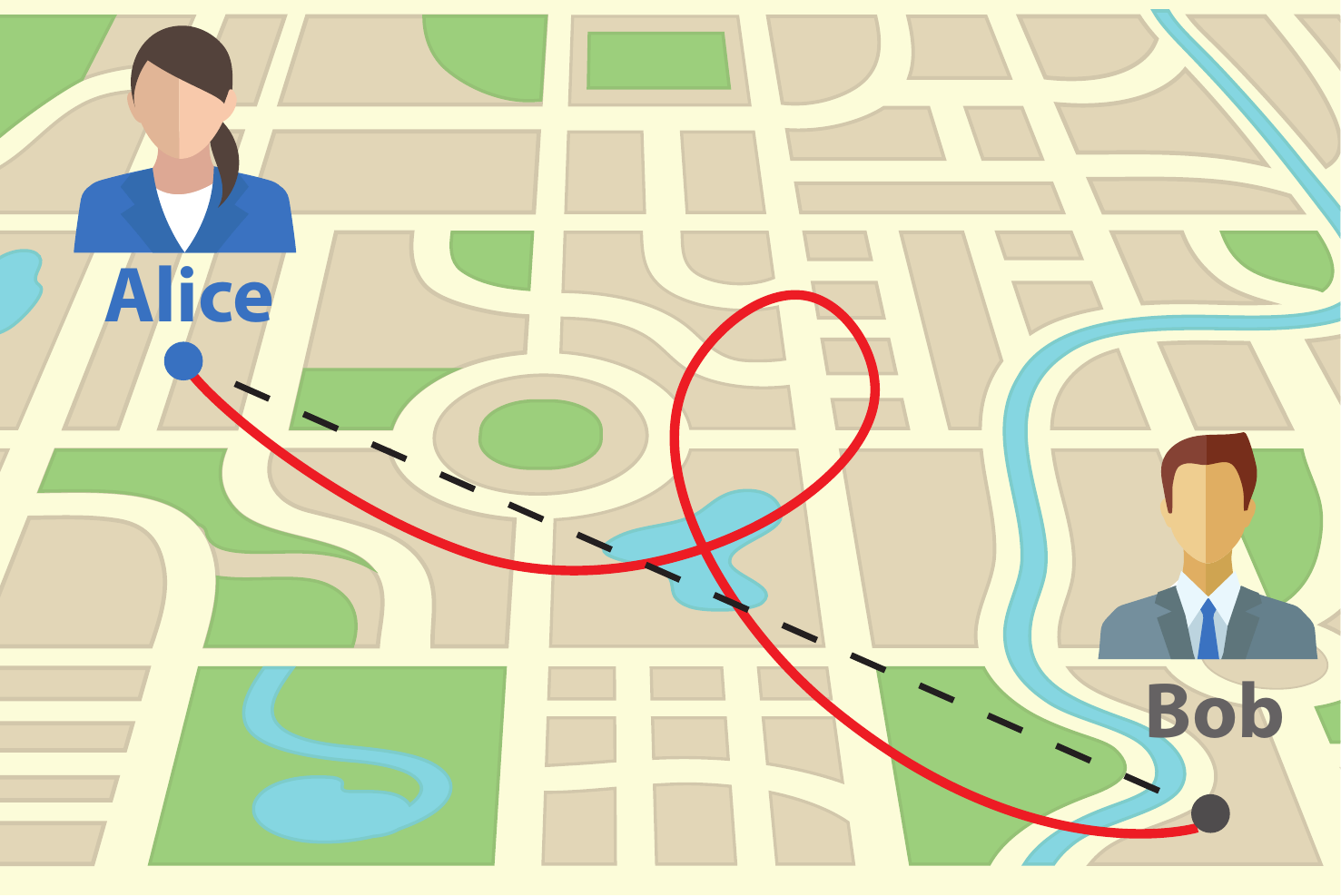} 
    \caption{Alice and Bob are connected by a noisy channel (e.g., an optical fiber) with unknown model of quantum disturbance.}
    \label{fig:AliceBob} 
    \vspace{4ex}
  \end{subfigure}
  \hfill
  \begin{subfigure}[t]{0.47\textwidth}
    \centering
    \includegraphics[width=.9\linewidth]{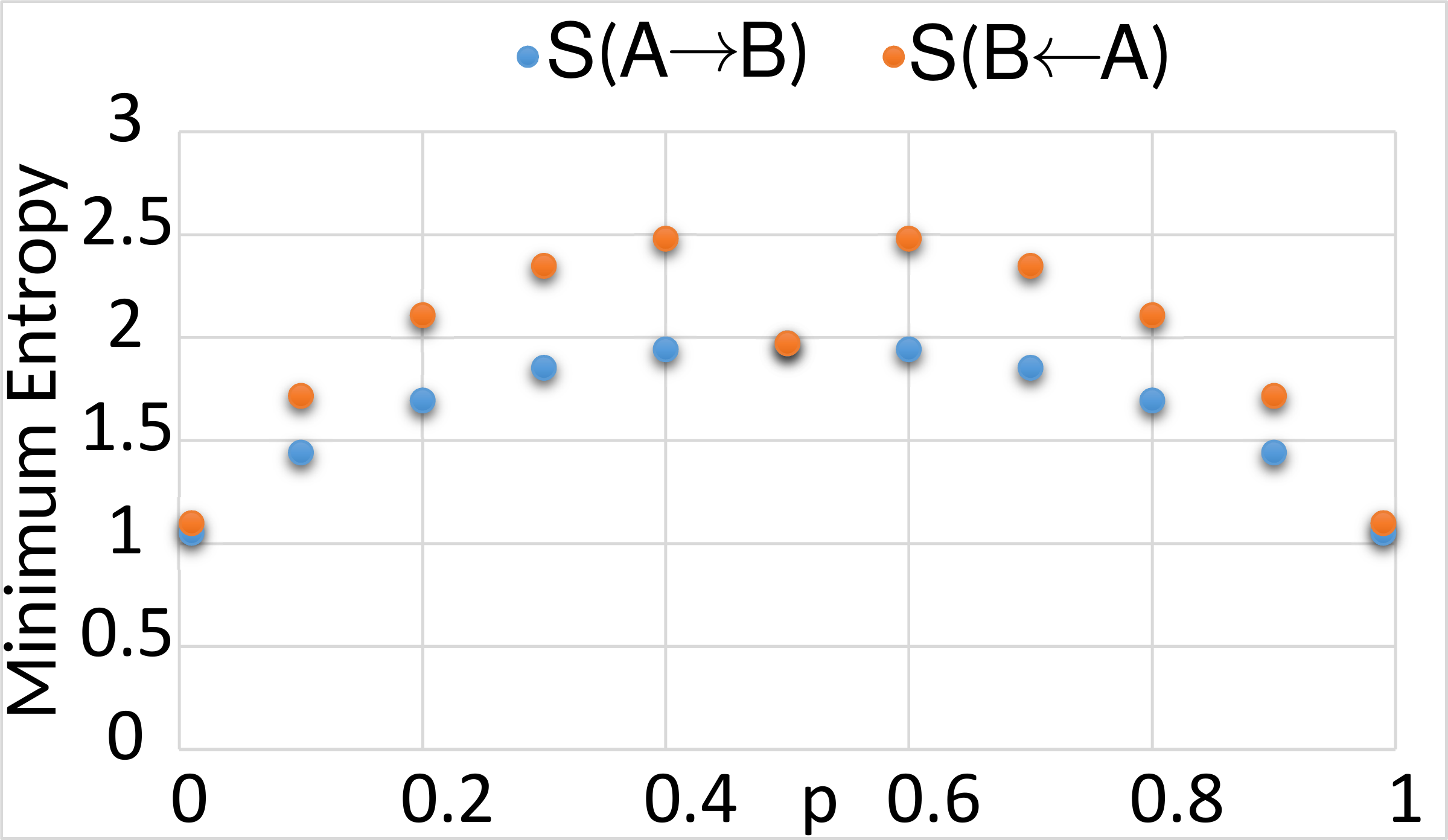}
	\caption{$S(A\to B)$ vs $S(A\gets B)$ for quantum symmetric channels with $q=0.4$, $0<p<1$.}
	\label{fig:qscPlot} 
    \vspace{4ex}
  \end{subfigure} 
  \begin{subfigure}[b]{0.47\textwidth}
    \centering
    \includegraphics[width=.95\linewidth]{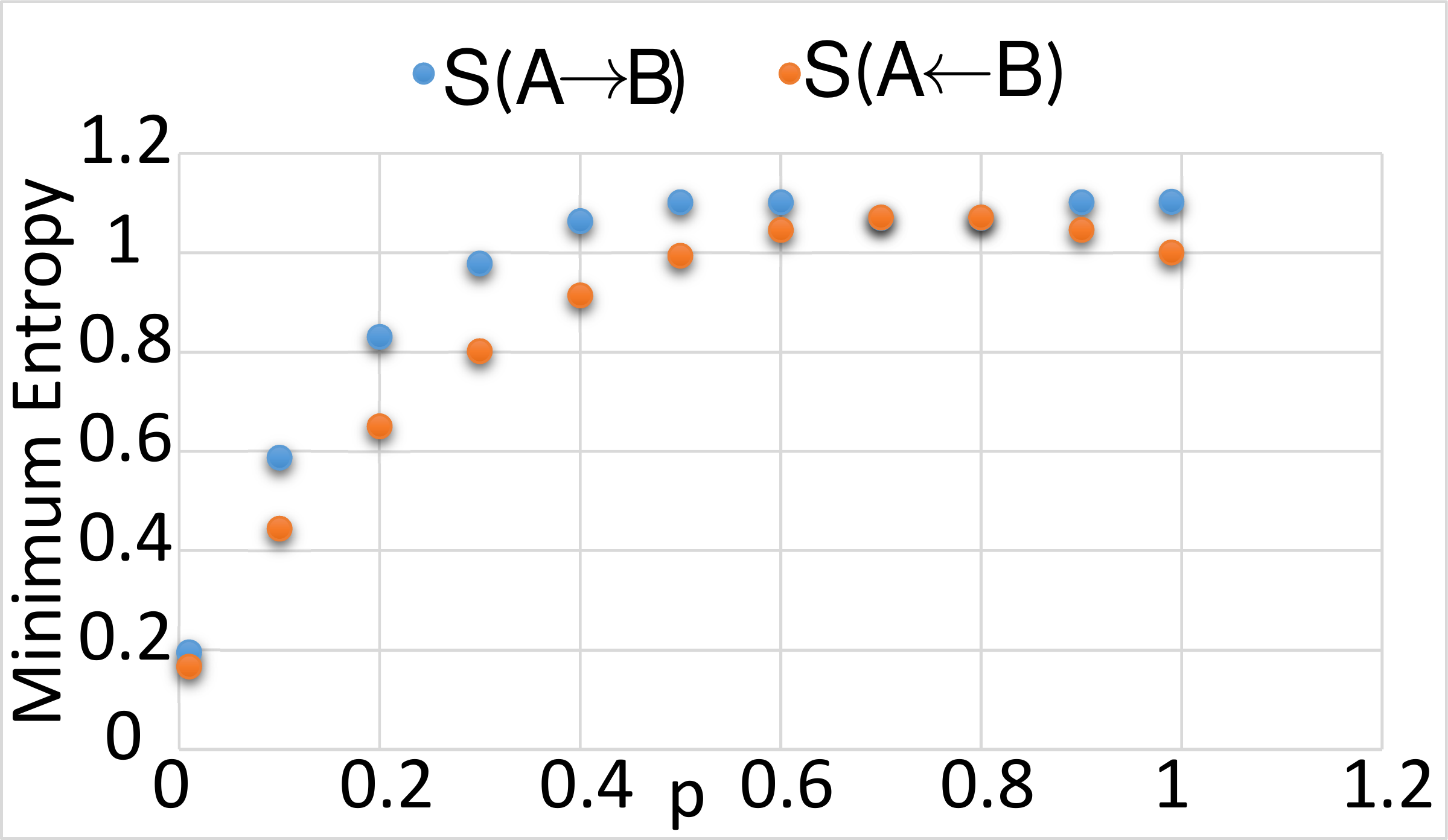}
        \caption{$S(A\to B)$ vs $S(A\gets B)$ for depolarization channels with joint density matrix $\rho_{AB}=0.4*\rho_{AB}^{0.6,0.8}+0.6*\rho_{AB}^{1/\sqrt{2},1/\sqrt{2}}$ is given for $0<p<1$.}
        \label{fig:counterex}
  \end{subfigure}
  \hfill
  \begin{subfigure}[b]{0.47\textwidth}
    \centering
    \includegraphics[trim = .7in 1.2in .7in 1.1in, clip, width=.9\linewidth]{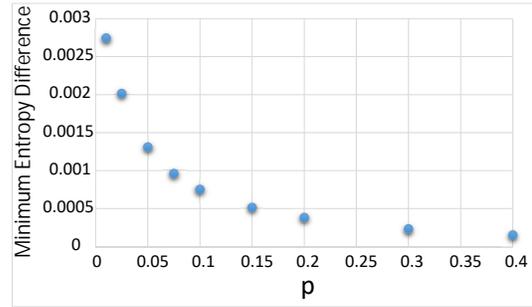}
    \caption{$\Delta (S)=S(A\gets B)-S(A\to B)$ for depolarization channels with joint density matrix $\rho_{AB}=0.4*\rho_{AB}^{0.6,0.8}+0.6*\rho_{AB}^{1/\sqrt{2},1/\sqrt{2}}$ is given for $0<p<1$.}
    \label{fig:depolarazation}
  \end{subfigure} 
  \caption{We use our model of quantum entropic causal inference on a synthetic dataset. Our goal is to find the originator of the message on a noisy channel when the noise model is unknown. Our approach successfully identifies the message originator. Here,  $S(A\to B)<S(A\gets B)$ indicates the originator of the message in the desired direction, i.e., $A$ to $B$.}
  \label{fig7} 
\end{figure}

Since there is no quantum cause-effect repository to verify the validity of our proposed algorithm, we put forward an experimental scheme that can be used to confront our theoretical framework. We consider a minimalistic model of an unknown message (possibly encrypted) with unknown origin in a two-node quantum network. The two nodes are connected by a noisy channel (e.g., an optical fiber) with unknown model of quantum disturbance, as depicted in Figure \ref{fig:AliceBob}. Consider a perfectly entangled system 
\begin{equation*}
    \rho_{AB} = \sum_i p_i |a_i\rangle|b_i\rangle(|a_i\rangle|b_i\rangle)^\dagger,
\end{equation*}
for some $p_i$ such that $\sum_i p_i=1$ and $p_i>0$. Alice prepares the two qubits ($\rho_{AB}$), and then communicates the second qubit  (quantum system $B$) to Bob. During the communication, there is a quantum noise model, which impacts the transmitted sub-system. We note that noise can arise in any quantum computation \citep{aggarwal2010volume}, and is not just limited to communications. In the presence of noise, we aim to find the originator of the message even though the noise model is unknown. More precisely we find if the originator of the message is Alice or Bob. In order to answer this question, we use the concept of causal inference. As mentioned in the structural example of Section III, we expect the qubit at the originator of the message to have causal relation to the qubit at the destination due to the unknown noise. Thus, we validate whether in this system $A$ will be the cause of $B$, to indicate that Alice is the originator of the message (or Alice prepared the 2 qubits and sent one to Bob).


We prove that only using the joint density matrix, we can identify the originator of the message. To verify the validation of \qeci, we use realistic quantum noisy links such as quantum symmetric channel and depolarizing channel (valid for quantum networking and quantum communications). Our work can lay the foundations of identifying originators of malicious activity on multi-node quantum networks.


In Appendix, we describe the Quantum Symmetric Channel (QSC), which is a quantum version of Binary Symmetric Channel. The proposed algorithm is run step-by-step to illustrate the approach. 
We  consider a generalization of QSC in the example below. In this example, the phase of the qubit is reversed with certain probability. 

We first consider the generalized Quantum Symmetric Channel.
Assume that $|a_1\rangle=|b_1\rangle=\frac{1}{\sqrt{2}} (|0\rangle +|1\rangle)$, and $|a_2\rangle=|b_2\rangle=\frac{1}{\sqrt{2}} (|0\rangle -|1\rangle)$. Let the quantum system be in the mixed state $|a_1\rangle |b_1\rangle$ with probability $q$, and $|a_2\rangle |b_2\rangle$ with probability $1-q$. We consider a generalization of Quantum Symmetric Channel in the following model, in which the phase of the qubit is reversed with certain probability.  The second qubit is transmitted over a quantum symmetric channels with error probability $p$, and is labeled $B$. Thus, the joint density matrix of $A,B$, $\rho_{AB}$,  can  be  written  as follows:
\begin{equation*}
\begin{split}
    \rho_{AB}&=
    (|a_1\rangle |b_1\rangle)(|a_1\rangle |b_1\rangle)^\dagger * q(1-p)+\\
    &(|a_1\rangle |b_2\rangle)(|a_1\rangle |b_2\rangle)^\dagger * qp +\\
    &(|a_2\rangle |b_1\rangle)(|a_2\rangle |b_1\rangle)^\dagger * (1-q)p +\\
    &(|a_2\rangle |b_2\rangle)(|a_2\rangle |b_2\rangle)^\dagger * (1-q)(1-p)
\end{split}
\end{equation*}
Since the $B$ subsystem was impacted by noise, we will now validate that system $A$ is the originator by verifying that $A$ is the cause of $B$ in this scenario, in a sense that Alice is the originator of the message (or Alice prepared the 2 qubits and sent one to Bob). 
To verify this using our proposed method \qeci, assume that the joint density matrix $\rho_{AB}$ is given with  $0<p<1$, and $q=0.4$. 

Results are shown in Figure \ref{fig:qscPlot}. Note that when $p=0.5$, $S(A\to B)=S(A\gets B)\simeq 1.97$, which indicates that $A$ and $B$ are uncorrelated in this case, as we expected.  Further, for $p=0$ and $p=1$, $A$ and $B$ are the symmetric and thus the originator of the message is unknown in these cases. Thus, we see that other than $p=0, 0.5, 1$, $S(A\to B)<S(A\gets B)$ indicating the originator of the message in the desired direction.

We now discuss the causal inference with a Depolarizing Channel
Assume that there are real numbers $\gamma_1$, $\gamma_2$, $\lambda_1$, and $\lambda_2$ such that $\gamma_1^2+\lambda_1^2=1$ and $\gamma_2^2+\lambda_2^2=1$. We consider a joint entangled system (of two qubits) as follows:
\begin{equation*}
    \begin{split}
        \rho_{AB}&=(\gamma_1^2|00\rangle+\gamma_1\lambda_1|01\rangle+\gamma_1\lambda_1|10\rangle+\lambda_1^2|11\rangle)(\gamma_1^2|00\rangle+\gamma_1\lambda_1|01\rangle+\gamma_1\lambda_1|10\rangle+\lambda_1^2|11\rangle)^\dagger * q+\\
        &(\gamma_2^2|00\rangle+\gamma_2\lambda_2|01\rangle+\gamma_2\lambda_2|10\rangle+\lambda_2^2|11\rangle)(\gamma_2^2|00\rangle+\gamma_2\lambda_2|01\rangle+\gamma_2\lambda_2|10\rangle+\lambda_2^2|11\rangle)^\dagger * (1-q)
    \end{split}
\end{equation*}

The system is a mixture of two pure density matrices. This quantum system has entanglement among the two quantum bits. Let the second quantum bit is transmitted over a \emph{quantum depolarizing channel}  with error probability $p$ \citep{nielsen2002quantum}. With this setup, the joint density matrix is given as $\rho_{AB} = q \rho_{AB}^{\gamma_1,\lambda_1} + (1-q) \rho_{AB}^{\gamma_2,\lambda_2}$, where $\rho_{AB}^{\gamma,\lambda}$ is given as follows:
\begin{equation*}
\begin{split}
    \rho_{AB}^{\gamma,\lambda}&=(\gamma^2|00\rangle+\gamma\lambda|01\rangle+\gamma\lambda|10\rangle+\lambda^2|11\rangle)(\gamma^2|00\rangle+\gamma\lambda|01\rangle+\gamma\lambda|10\rangle+\lambda^2|11\rangle)^\dagger * (1-p)+\\
		&(\gamma^2|00\rangle-\gamma\lambda|01\rangle+\gamma\lambda|10\rangle-\lambda^2|11\rangle)(\gamma^2|00\rangle-\gamma\lambda|01\rangle+\gamma\lambda|10\rangle-\lambda^2|11\rangle)^\dagger  * (p/3)+\\
		&(\gamma\lambda|00\rangle+\gamma^2|01\rangle+\lambda^2|10\rangle+\gamma\lambda|11\rangle)(\gamma\lambda|00\rangle+\gamma^2|01\rangle+\lambda^2|10\rangle+\gamma\lambda|11\rangle)^\dagger  * (p/3)+\\
		&(-\gamma\lambda|00\rangle+\gamma^2|01\rangle-\lambda^2|10\rangle+\gamma\lambda|11\rangle)(-\gamma\lambda|00\rangle+\gamma^2|01\rangle-\lambda^2|10\rangle+\gamma\lambda|11\rangle)^\dagger  * (p/3)
\end{split}
\end{equation*}
We note that $A$ and $B$ coexist in the quantum system, and thus the joint density matrix has been obtained. We already know that $A$ is the originator of the qubits, and thus expect  $A\to B$ as the corresponding directed graph.
Figure \ref{fig:depolarazation} shows the $\Delta (S)=S(A\gets B)-S(A\to B)$ for different probability errors $0<p<1$, using \qeci. 

\textbf{Why Should We Not Map Quantum to Classical Directly?}
Here, we show  why classical entropic causal inference do not directly apply to the quantum case. 
We emphasize that although a joint density operator (matrix) can be converted to a joint probability distribution (as explained in the following example), we lose some quantum information due to the loss of entanglement. The following example shows that converting a joint density matrix $\rho_{AB}$ directly to a joint probability distribution $p(A,B)$, and then applying classical entropic causal inference on $p(A,B)$ will not lead to the correct results.

Here, we show a counter example.
Assume the depolarizing channel as described in this section. We already know that $A$ causes $B$ in this model.  To convert the joint density matrix $\rho_{AB}$, we use a rotational procedure explained as follows: Assume that $\rho_{AB}$ is rotated using a unitary matrix $U$. Let us say $\rho_{AB} = U \rho'_{AB} U^\dagger$. So, the joint density matrix $\rho'_{AB}$ is computed as $\rho'_{AB}= U^\dagger \rho_{AB}U$. To compute the unitary matrix $U$ for a given $\rho_{AB}$ we use the eigenspaces of $\rho_A$ and $\rho_B$, where $\rho_A=\textbf{\textrm{Tr}}_Y(\rho_{AB})$ and $\rho_B=\textbf{\textrm{Tr}}_X(\rho_{AB})$ are computed by tracing out $B$ and $A$, respectively. This simple observation enables us to design a procedure that converts a joint density matrix $\rho_{AB}$ to a joint probability distribution $p(A,B)$ in a way that it takes into account the rotation. This procedure is formally described in Algorithm \ref{alg:rotate}. 

    By converting the joint density matrix $\rho_{AB}$ directly to a joint probability distribution $p(A,B)$, using Algorithm \ref{alg:rotate}, and then applying classical entropic causal inference, i.e., Algorithm \ref{alg:ceci} on $p(A,B)$ we obtain the results represented in Figure \ref{fig:counterex} which is opposite to the expected  direction in all cases.

\section{Conclusions}

This paper provides a novel approach for quantum entropic causal inference. As a part of the approach, an algorithmic greedy solution is provided for the minimum-entropy quantum marginal problem and a notion of  instance conditional density matrices is developed. The approach is validated on quantum noisy link, where the approach detects the expected causal relation. 

We note that the joint density matrix required in our analysis can be estimated from measurements using quantum tomography \citep{d2003quantum}, and is beyond the scope of this paper. The extension of the problem to general quantum causality  graph relations between multiple variables is an open problem for the future. 

\section*{Acknowledgements}
The authors would like to thank Wenbo Sun for his comments on the definition of instance conditional density matrices and Xueji Wang for generating Figures \ref{fig:AliceBob}.  This research was supported by the Defense Advanced Research Projects Agency (DARPA)  Quantum Causality [Grant No. HR00112010008].
\bibliographystyle{IEEEtran}
\bibliography{reference}
\clearpage
\appendices
\renewcommand{\appendixname}{Supplementary Information}
\renewcommand{\thesection}{Supplementary Information \arabic{section}}
\section{Classical Entopic Causal Inference}\label{sec:classic}

Consider that the  joint distribution $P(X,Y)$ between two observed variables is given. The aim of the entropic causal inference is to determine  the causality relation between $X$ and $Y$. The authors of \citep{Murat2017} proposed an approach to determine the causal direction by assuming that minimization of the information entropy of exogenous variable (i.e., one that is not caused by any others in the model) identifies the causal direction. The key assumption used for determining the causal direction in  \citep{Murat2017} is given as follows. 

\begin{assumption}\label{mainassump}
Entropy of the exogenous variable $E$ is small
in the true causal direction. In other words, the exogenous variable has
lower entropy in the true causal direction than the entropy in the wrong direction.
\end{assumption}

However,  in the causal graph $X\to Y$, where $Y = f(X, E)$, finding the exogenous variable $E \perp\!\!\!\perp X$ (i.e., $E$ is marginally independent of $X$) with minimum
entropy is an NP hard problem due to the following argument. Assume that $Y = f(X, E)$ and $E \perp\!\!\!\perp X$. Define the function $f_x$ as
$\left\{\begin{array}{l}
     f_x:\mathcal{E}\to \mathcal{Y}  \\
     f_x(E)=f(x,E) 
\end{array}\right.$. Using this definition and the fact that $E \perp\!\!\!\perp X$, we have:
\begin{align*}
&p(Y = y|X = x) && \text{}\\
&= p(f_x(E) =y|X = x) && \text{by the definition of $f_x$}\\
&= p(f_x(E) =y) && \text{because $E \perp\!\!\!\perp X$}
\end{align*}

This means that the conditional distributions $p(Y|X = x)$ can be considered and treated as distributions $p(f_x(E))$ by applying the function $f_x$ on an exogenous (unobserved) variable $E$. In other words, the problem of finding the exogenous variable $E$ with minimum entropy given the joint distribution $p(X,Y)$ is
equivalent to the problem of finding the minimum entropy joint distribution of the random variables $U_i = (Y |X = x_i)$, given the marginal distributions $p(Y |X = x_i)$ \citep{Murat2017}. However, the problem of minimizing entropy subject to marginal constraints is non-convex and NP hard 
\citep{kovavcevic2015entropy}, and therefore it probably requires exponential time in the worst case. Importantly, however, the minimum entropy coupling problem can be tackled effectively using a heuristic search, as shown in \citep{kocaoglu2017entropic,cicalese2017find,rossi2019greedy}. 

The detailed algorithm for determining causal direction in \citep{Murat2017} is given in Algorithm \ref{alg:ceci} that uses Algorithm \ref{alg:jem} for joint entropy minimization.  The key steps in the algorithm include:
\begin{enumerate}
	\item Computing marginal and conditional probability distributions from $p(X,Y)$: In this step, we compute $p(X|Y)$ and $p(Y|X)$. 
	\item Compute minimum entropy of the exogenous variables from $X$ to $Y$ and from $Y$ to $X$: This is where the problem of minimizing entropy subject to marginal constraints is used, and a greedy algorithm is provided for this problem. The greedy algorithm is provided in Algorithm \ref{alg:jem}. This algorithm has been shown to always finds a local minimum, which 
	is within an additive guaranteed gap from the unknown global optimum \citep{kocaoglu2017entropic}.
	\item Compare the minimum entropy from $X$ to $Y$ and from $Y$ to $X$, and return the causal model with smaller entropy. 
\end{enumerate}

The second step finds the minimimum joint entropy subject to marginal constraints, which we refer to as the  minimum entropy couplings problem. This problem has a pivotal role in entropic causal inference, we take a look at this problem from quantum computing point of view in Section \ref{sec:qcoupling}.


\begin{algorithm}[!ht]
	\caption{Classical Entropic Causal Inference \citep{Murat2017}}\label{alg:ceci}
	\SetAlgoLined
	\KwIn{The joint probability distribution $p(X,Y)$ for a pair of random variables $X$ and $Y$.}
	\KwOut{The cause-effect relationship between $X$ and $Y$.}
	\tcc{\textcolor{red}{Step 1: Compute marginal and conditional probability distributions from $p(X,Y).$}}
	\tikzmk{A}
	Compute $p(X)$ and $p(Y)$\;
	Compute $p(X|Y)$ and $p(Y|X)$\;
	\nonl\tikzmk{B}\boxit{green!50}
	\tcc{\textcolor{red}{Step 2: Compute minimum entropy of the exogenous variables from $X$ to $Y$ and from $Y$ to $X$.}}
	\tcc{\textcolor{blue}{Estimation of minimum entropy $H(E)$ for the causal model $Y = f(X, E).$}}
	\nl\tikzmk{A}
	$M=$ the columns of the conditional probability distribution $p(Y|X)$\;
	$H(E)$=\textbf{JointEntropyMinimization}($M$)\;
	$H(A\to B)=H(X)+H(E)$\;
	\tcc{$H(.)$ is the Shannon entropy i.e., $H(X)=-\sum_{i=1}^np(X_i)\log(p(X_i))$}
	\nonl\tikzmk{B}\boxit{blue!35}
	\tcc{\textcolor{blue}{Estimation of minimum entropy $H(E')$ for the causal model $X = g(Y, E').$}}
	\nl\tikzmk{A}
	$M=$ the columns of the conditional probability distribution $p(X|Y)$\;
	$H(E')$=\textbf{JointEntropyMinimization}($M$)\;
	\tcc{Use Algorithm \ref{alg:jem}: Joint Entropy Minimization Algorithm}
	$H(A\gets B)=H(Y)+H(E')$\;
	\nonl\tikzmk{B}\boxit{blue!35}
	\tcc{\textcolor{red}{Step 3: Compare the minimum entropy from $X$ to $Y$ and from $Y$ to $X$, and return the causal model with smaller entropy.}}
	\nl\tikzmk{A}
	\uIf{($H(A\to B)<H(A\gets B)$)}{\textbf{return} The causal model is of the form $A\to B$.}\Else{\textbf{return} The causal model is of the form $A\gets B$.}
	\nonl\tikzmk{B}\boxit{red!35}
\end{algorithm}
\begin{algorithm}[!ht]
	\caption{\textbf{JointEntropyMinimization}: Joint Entropy Minimization Algorithm \citep{Murat2017}}\label{alg:jem}
	\SetAlgoLined
	\KwIn{Marginal distributions of $m$ variables each with $n$ states,
		in matrix form $M =\begin{bmatrix}
			p_1^T;\cdots;p_m^T\\
		\end{bmatrix}$.}
	\KwOut{Minimum entropy of the joint distributions.}
	$e = [\quad ]$\;
	Sort each row of $M$ in decreasing order\;
	Find minimum of maximum of each row: $r\gets \min_i(p_i^T(1))$\;
	\While{$r>0$}{
		$e\gets [e,r]$\;
		Update maximum of each row: $p_i^T(1)\gets p_i^T(1)-r$, $\forall i$\;
		Sort each row of M in decreasing order\;
		$r\gets \min_i(p_i^T(1))$
	}
	\textbf{return} $Entropy(e)=-\sum_ie_i\log(e_i)$.
\end{algorithm}

\section{Pseudocode of Main Algorithms}\label{sec:pseudocode}
Algorithm \ref{alg:qgreedy} lists the pseudo-code of our proposed algorithm for solving the minimum-entropy quantum marginal problem, discussed in section \ref{sec:greedyalg}.
\begin{algorithm}[!ht]
\caption{\textbf{QJointEntropyMinimization}: Greedy Entropy Minimization Algorithm for Quantum Marginal}\label{alg:qgreedy}
    \SetAlgoLined
	\KwIn{Density matrices of quantum systems ${U_1}$ and ${U_2}$, i.e., $\rho_{U_1}$ and $\rho_{U_2}$ of the size $m$-by-$m$ and $n$-by-$n$, respectively.}
	\KwOut{Minimum entropy $S(\rho_E)$ of the joint density matrix.}
	\tcc{\textcolor{blue}{\scriptsize{Step 1:Compute eigenvalues of $\rho_{U_1}$ and $\rho_{U_2}$}}}
	$p_{U_1}=eig(\rho_{U_1})$\;
	$q_{U_2}=eig(\rho_{U_2})$\;
	\tcc{\textcolor{blue}{\scriptsize{Step 2: Apply Joint Entropy Minimization Algorithm \citep{kocaoglu2017entropic} on $p_{U_1}$ and $q_{U_2}$.}}}
	$e = [\quad]$\;
	Initialize the matrix $M_{ij} = 0, i=1,\dots, m, j = 1,\dots, n$\;
	Initialize $r = 1$\;
	\While{$r>0$}{
	e=[e,r]\;
	($\{p_{U_1}, q_{U_2}\}, r$) = \textbf{UpdateRoutine}($\{p_{U_1}, q_{U_2}\}, r$)}
	\textbf{UpdateRoutine}($\{p_{U_1}, q_{U_2}\}, r$)\{
	
	$i = \textrm{argmax}_k\{p_k\}$\;
	$j = \textrm{argmax}_k\{q_k\}$\;
	$M_{ij} = \min \{p_i, q_j\}$\;
	$p_i = p_i - M_{ij}$\;
	$q_i = q_i - M_{ij}$\;
	$r = r - M_{ij}$\;
	\}
	
	$S(\rho_E)=-\sum_ke_k\log(e_k)$\;
	\textbf{return} $S(\rho_E)$.
\end{algorithm}
\newpage
Algorithm \ref{alg:qeci} lists the pseudo-code of our proposed algorithm for quantum entropic causal inference problem, discussed in section \ref{sec:qeci}.
\begin{algorithm}[!ht]
\caption{\qeci: \footnotesize{Quantum Entropic Causal Inference}}\label{alg:qeci}
    \SetAlgoLined
	\footnotesize\KwIn{The joint density matrix $\rho_{AB}$ for a pair of quantum systems $A$ and $B$.}
	\KwOut{The cause-effect relationship between $A$ and $B$.}
	\tcc{\textcolor{red}{Phase 1: Compute reduced density matrices $\rho_A$ and $\rho_B$ as well as density matrix $\rho_{BA}$ from $\rho_{AB}.$}}
	\tikzmk{A}
	$\rho_A=\Tr_B(\rho_{AB})$\;
	$\rho_B=\Tr_A(\rho_{AB})$\;
	Compute $\rho_{BA}$ by reordering the entries of $\rho_{AB}$, appropriately.\;
	\nonl\tikzmk{B}\boxit{green!50}
	\tcc{\textcolor{blue}{Estimation of minimum entropy $S(A\to B)$ for the causal model $B = f(A, E).$}}
	\tikzmk{A}
	\tcc{\textcolor{red}{Phase 2a: Compute eigenvalues and eigenvectors of $\rho_A.$}}
	\nl$[V, D] = eig(\rho_A)$\;
	\tcc{Computes diagonal matrix $D$ of eigenvalues and matrix $V$ whose columns are the corresponding eigenvectors, so that $\rho_A*V = V*D$.}
	\For{($d_i\in D$)}{
	\tcc{\textcolor{blue}{Compute instance conditional density matrices.}}
	    $|a_i\rangle\langle a_i|=e_i*e_i^T$\;
	    \tcc{$e_i$ is the $i$'th column of $V$, and $e_i^T$ is the transpose of $e_i$.}
	    $\rho_i=\Tr_A(\rho_{AB}\star|a_i\rangle\langle a_i|)$\;
	    \tcc{$I$ is the identity matrix.}
	    $\rho_{B||a_i\rangle_A}=\rho_i/trace(\rho_i)$\;
	    \tcc{\textcolor{blue}{Compute marginal distributions.}}
	    $B_i=eig(\rho_{B||a_i\rangle_A})$\;
	    Add $B_i$ to the $i$'th row of the matrix $M$
	}
	\tcc{\textcolor{red}{Phase 3a: Apply Algorithm \ref{alg:qgreedy} to estimate joint entropy minimization.}}
	$S(\rho_E)\approx$\textbf{ QJointEntropyMinimization}($M$)\;
	$S(A\to B)=S(\rho_A)+S(\rho_E)$\;
	\nonl\tikzmk{B}\boxit{blue!35}
	\tcc{\textcolor{blue}{Estimation of minimum entropy $S(A\gets B)$ for the causal model $A = g(B, E').$}}
	\tikzmk{A}
	\tcc{\textcolor{red}{Phase 2b: Compute eigenvalues and eigenvectors of $\rho_B.$}}
	\nl$[V, D] = eig(\rho_B)$\;
	\For{($d_i\in D$)}{
	\tcc{\textcolor{blue}{Compute instance conditional density matrices.}}
	    $|b_i\rangle\langle b_i|=e_i*e_i^T$\;
	    $\rho_i=\Tr_B(\rho_{BA}\star|b_i\rangle\langle b_i|)$\;
	    $\rho_{A||b_i\rangle_B}=\rho_i/trace(\rho_i)$\;
	    \tcc{\textcolor{blue}{Compute marginal distributions.}}
	    $B_i=eig(\rho_{A||b_i\rangle_B})$\;
	    Add $B_i$ to the $i$'th row of the matrix $M$\;
	}
	\tcc{\textcolor{red}{Phase 3b: Apply Algorithm \ref{alg:qgreedy} to estimate joint entropy minimization.}}
	$S(\rho_{E'})\approx$\textbf{ QJointEntropyMinimization}($M$)\;
	$S(A\gets B)=S(\rho_B)+S(\rho_{E'})$\;
	\nonl\tikzmk{B}\boxit{blue!35}
	\tcc{\textcolor{red}{Phase 4: Compare the minimum entropy from $A$ to $B$ and from $B$ to $A$.}}
	\nl\tikzmk{A}
	\uIf{($S(A\to B)<S(A\gets B)$)}{\textbf{return} The causal model is of the form $A\to B$.}\Else{\textbf{return} The causal model is of the form $A\gets B$.}
	\nonl\tikzmk{B}\boxit{red!35}
\end{algorithm}

Algorithm \ref{alg:rotate} lists the pseudo-code for a procedure that converts a joint density matrix $\rho_{AB}$ to a joint probability distribution $p(A,B)$ in a way that it takes into account the rotation, as discussed in section \ref{sec:examples}.
\begin{algorithm}[!ht]
\caption{Rotational procedure for computing the joint probability distribution of a joint density matrix}\label{alg:rotate}
    \SetAlgoLined
	\KwIn{Joint density matrix of quantum systems $A$ and $B$ i.e., $\rho_{AB}$.}
	\KwOut{Joint probability distribution $p(A,B)$ corresponding to the joint density matrix $\rho_{AB}$.}
	\tcc{\textcolor{blue}{\scriptsize{Compute eigenvalues and eigenvectors of $\rho_A$}}}
	$[V_1, D_1] = eig(\rho_A)$\;
	\tcc{\textcolor{blue}{\scriptsize{Compute eigenvalues and eigenvectors of $\rho_B$}}}
	$[V_2, D_2] = eig(\rho_B)$\;
	\tcc{\textcolor{blue}{\scriptsize{Rotational procedure}}}
	$U= V_1\otimes V_2$\;
	$\rho'_{AB}= U^\dagger \rho_{AB}U$\;
	\textbf{return} $p(A,B)$ as the entries on the main diagonal of $\rho'_{AB}$.
\end{algorithm}

\section{Discussions}\label{sec:discussion}
\subsection{Complexity Analysis of \qeci.} First, we discuss the time complexity of \qeci. Assume that $\rho_A$ and $\rho_B$ of the size $m$-by-$m$ and $n$-by-$n$, respectively. The most expensive parts of \qeci~ are the computation of eigenvalues and eigenvectors as well as joint entropy minimization algorithm. So, we have:
\begin{itemize}[noitemsep]
	\item To compute eigenvalues and eigenvectors of $\rho_A$, i.e., $[V, D] = eig(\rho_A)$, in the worst case scenario the time complexity is $O(m^3)$ \citep{press1988numerical}.
	\item To compute $B_i=eig(\rho_{B||a_i\rangle_A})$ ($m$ times, line 5-11) the time complexity is $O(m^2\log(m))$ because for symmetric tridiagonal eigenvalue problems all eigenvalues (without eigenvectors) can be computed numerically in time $O(m\log(m))$, using bisection on the characteristic polynomial \citep{coakley2013fast}.
	\item To apply joint entropy minimization algorithm (Algorithm \ref{alg:jem}) in line 12, the time complexity is $O(m^4\log(m))$, it can easily be reduced to $O(m^2\log(m))$ by dropping the sorting step \citep{Murat2017}.
	\item To compute eigenvalues and eigenvectors of $\rho_B$, i.e., $[V, D] = eig(\rho_B)$, in the worst case scenario the time complexity is $O(n^3)$ \citep{press1988numerical}.
	\item To compute $B_i=eig(\rho_{A||b_i\rangle_B})$ ($m$ times, line 15-21) the time complexity is $O(n^2\log(n))$.
	\item To apply joint entropy minimization algorithm (Algorithm \ref{alg:jem}) in line 22, the time complexity is $O(n^4\log(n))$, it can easily be reduced to $O(n^2\log(n))$ by dropping the sorting step \citep{Murat2017}.
\end{itemize}
As a result the time complexity of \qeci~is $O(\max(m^3,n^3))$. It is not difficult to see that the space complexity of \qeci~is $O(m^2n^2)$ due to the size of $\rho_{AB}$ and $\rho_{BA}$, i.e., $mn$-by-$mn$.

From the viewpoint of computational complexity, a concave minimization problem (e.g., Minimum Entropy Quantum Coupling Problem) is NP-hard \citep{horst2013global}. Assume that $\rho_A$ and $\rho_B$ are of the size $m$-by-$m$ and $m$-by-$m$, respectively. So, $\rho_{AB}$ is of the size $mn$-by-$mn$, and all of $m$ instance conditional density matrices $\rho_{B||a_i\rangle}, i=1,\cdots,m$ are of the size $m$-by-$m$. This indicates that the solution of the minimum-entropy quantum marginal problem: $\min S(\rho_{B||a_1\rangle,\cdots,B||a_{m}\rangle})$
is a matrix of the size $m^{m}$-by-$m^{m}$. Similarly, for the minimum-entropy quantum marginal problem: $\min S(\rho_{A||b_1\rangle,\cdots,A||b_{n}\rangle})$
is a matrix of the size $n^{n}$-by-$n^{n}$. So, the total space complexity of the optimization approach is $O(\max(m^m,n^n))$. As a result this approach is impractical.

\subsection{Rotational Invariance of \qeci}

Let us assume that $\rho_A$ is rotated using a unitary matrix $U$. Let us say $\rho_Z = U \rho_A U^\dagger$. Then, the causal direction between $A$ and $B$ is the same as that between $Z$ and $Y$. In order to see this, we note that the second phase of the \qeci\  will result in $|z_i\rangle = U |x_i\rangle$. Further, we note that from the definition of conditional densities, $\rho_{Z|Y=|y_j\rangle} = U\rho_{X|Y=|y_j\rangle}U^\dagger$ and $\rho_{Y|X=|x_i\rangle} = \rho_{Y|Z=U|x_i\rangle}$. Further, up to a rotation of the eigenstates in $\rho_{Z|Y=|y_j\rangle}$, the eigenvalues required for quantum marginal problem remain the same. Since these eigenvalues remain the same, the entropy of the quantum coupling remains the same. Since the entropy of the quantum density matrix is independent of rotations, the overall $S(X\to Y)$ and $S(Y\to X)$ are the same as $S(Z\to Y)$ and $S(Y\to Z)$, respectively. By symmetry, the result also holds if $Y$ is rotated by a unitary matrix. 
\subsection{Classical vs Quantum Approach}\label{sec:ClassicvsQuantum}
We emphasize that
causal inference using entanglement entropy of hidden variables is a fundamentally new approach that also reduces to the classical causal inference method as expected. In other words, \qeci~is a unified framework
for classical and quantum causal inference. In fact, quantum entropic causal inference approach is a generalization of the classical entropic causal inference because any classical probability distribution $p:X \to [0,1]$ can be written as a density matrix $\rho_A$ by writing down a matrix
with the probabilities along its diagonal. In fact, it does so in more than one way. However, for our purpose the diagonal approach is enough.  For example, consider the following joint probability distribution $p(X,Y)$:
\begin{center}
    \begin{tabular}{c|c|c}
		& $y_1$   & $y_2$ \\
		\midrule
		$x_1$& $\frac{1}{16}$ & $\frac{3}{16}$ \\
		\midrule
		$x_2$& $\frac{5}{16}$ & $\frac{7}{16}$ \\
	\end{tabular}
\end{center}

\begin{figure*}[!ht]
	\centering
	\begin{minipage}{.5\textwidth}
		\begin{tikzpicture}
			\tikzset{font=\scriptsize}
			\tikzstyle{io} = [rectangle, rounded corners, text width=3cm, minimum height=1cm,text centered, draw=black, fill=red!30, inner sep=2pt]
			\tikzstyle{process} = [rectangle, text width=3cm, minimum height=1cm, text centered, draw=black, fill=orange!30, inner sep=2pt]
			\tikzstyle{decision} = [diamond, text width=1.8cm, minimum height=1cm, text centered, draw=black, fill=green!30, inner sep=2pt]
			\tikzstyle{arrow} = [thick,->,>=stealth]
			
			\node (in1) [io] {Input: joint probability distribution $p(X,Y)$};
			\node (pro1) [process, below of=in1, yshift=-.5cm] {Compute marginal distributions $p(X)=\sum_{Y}p(X,Y)$ $p(Y)=\sum_Xp(X,Y)$};
			\node (pro2) [process, below left =1.5cm and -1cm  of pro1, yshift=1cm] {Compute conditional distribution $p(X|Y)=\frac{p(X,Y)}{p(Y)}$};
			\node (pro3) [process, below right =1.5cm and -1cm  of pro1, yshift=1cm] {Compute conditional distribution $p(Y|X)=\frac{p(X,Y)}{p(X)}$};
			\node (pro4) [process, below of = pro2, yshift=-.75cm] {Apply Algorithm \ref{alg:jem} on $M=p(Y|X)$ to estimate joint entropy minimization
				$H(E)$};
			\node (pro5) [process, below of = pro3, yshift=-.75cm] {Apply Algorithm \ref{alg:jem} on $M=p(Y|X)$ to estimate joint entropy minimization
				$H(E')$};
			\node (pro6) [process, below of = pro4, yshift=-.75cm] {$H(X\to Y)=H(X)+H(E)$};
			\node (pro7) [process, below of = pro5, yshift=-.75cm] {$H(X\gets Y)=H(Y)+H(E')$};
			\node (dec1) [decision, below of=pro1, yshift=-6.25cm] {$H(X\to Y)<H(X\gets Y) ?$};
			\node (io2) [io, below left =1.5cm and -.5cm  of dec1, yshift=.5cm] {Output: The causal model is of the form $X\to Y$.};
			\node (io3) [io, below right =1.5cm and -.5cm  of dec1, yshift=.5cm] {Output: The causal model is of the form $X\gets Y$.};
			
			\draw [arrow] (in1) -- (pro1);
			\draw [arrow] (pro1) -- (pro2);
			\draw [arrow] (pro1) -- (pro3);
			\draw [arrow] (pro2) -- (pro4);
			\draw [arrow] (pro3) -- (pro5);
			\draw [arrow] (pro4) -- (pro6);
			\draw [arrow] (pro5) -- (pro7);
			\draw [arrow] (pro6) -- (dec1);
			\draw [arrow] (pro7) -- (dec1);
			\draw [arrow] (dec1) -- node[anchor=east]  {yes}(io2);
			\draw [arrow] (dec1) -- node[anchor=west]  {no}(io3);
		\end{tikzpicture}
	\end{minipage}%
	\begin{minipage}{.5\textwidth}
		\begin{tikzpicture}
			\tikzset{font=\scriptsize}
			\tikzstyle{io} = [rectangle, rounded corners, text width=3cm, minimum height=1cm,text centered, draw=black, fill=red!30, inner sep=2pt]
			\tikzstyle{process} = [rectangle, text width=3.3cm, minimum height=1cm, text centered, draw=black, fill=orange!30, inner sep=2pt]
			\tikzstyle{decision} = [diamond, text width=1.8cm, minimum height=1cm, text centered, draw=black, fill=green!30, inner sep=2pt]
			\tikzstyle{arrow} = [thick,->,>=stealth]
			
			\node (in1) [io] {Input: joint density matrix $\rho_{AB}$};
			\node (pro1) [process, below of=in1, yshift=-.5cm] {Compute partial traces $\rho_A=\Tr_Y(\rho_{AB})$
				$\rho_B=\Tr_X(\rho_{AB})$};
			\node (pro2) [process, below left =1.5cm and -1cm  of pro1, yshift=1cm] {Compute eigenvalues and eigenvectors $[V,D]=eig(\rho_B)$};
			\node (pro3) [process, below right =1.5cm and -1cm  of pro1, yshift=1cm] {Compute eigenvalues and eigenvectors $[V,D]=eig(\rho_A)$};
			\node (pro4) [process, below of=pro2, yshift=-.95cm] {Compute instance conditional density matrices, $\forall d_i\in D:$
				$\rho_i=\Tr_B(|b_i\rangle\langle b_i|\star\rho_{BA})$
				$\rho_{A|B=|b_i\rangle}=\rho_i/trace(\rho_i)$};
			\node (pro5) [process, below of=pro3, yshift=-.95cm] {Compute instance conditional density matrices, $\forall d_i\in D:$
			
				$\rho_i=\Tr_A(|a_i\rangle\langle a_i|\star\rho_{AB})$
				$\rho_{B|A=|a_i\rangle}=\rho_i/trace(\rho_i)$};
			\node (pro6) [process, below of = pro4, yshift=-1.2cm] {Compute marginal distributions:\\
				$B_i=eig(\rho_{B|A=|a_i\rangle})$
				Add $B_i$ to the $i$'th row of the matrix $M$};
			\node (pro7) [process, below of = pro5, yshift=-1.2cm] {Compute marginal distributions:\\
				$B_i=eig(\rho_{A|B=|b_i\rangle})$
				Add $B_i$ to the $i$'th row of the matrix $M$};
			\node (pro8) [process, below of = pro6, yshift=-.85cm] {Apply Algorithm \ref{alg:qgreedy} on $M$ to estimate joint entropy minimization
				$S(\rho_E)$};
			\node (pro9) [process, below of = pro7, yshift=-.85cm] {Apply Algorithm \ref{alg:qgreedy} on $M$ to estimate joint entropy minimization
				$S(\rho_{E'})$};
			\node (pro10) [process, below of = pro8, yshift=-.5cm] {$S(A\to B)=S(\rho_A)+S(\rho_E)$};
			\node (pro11) [process, below of = pro9, yshift=-.5cm] {$S(A\gets B)=S(\rho_B)+S(\rho_{E'})$};
			\node (dec1) [decision, below of=pro1, yshift=-10cm] {$S(A\to B)<S(A\gets B) ?$};
			\node (io2) [io, below left =1.5cm and -.5cm  of dec1, yshift=.5cm] {Output: The causal model is of the form $A\to B$.};
			\node (io3) [io, below right =1.5cm and -.5cm  of dec1, yshift=.5cm] {Output: The causal model is of the form $A\gets B$.};
			
			\draw [arrow] (in1) -- (pro1);
			\draw [arrow] (pro1) -- (pro2);
			\draw [arrow] (pro1) -- (pro3);
			\draw [arrow] (pro2) -- (pro4);
			\draw [arrow] (pro3) -- (pro5);
			\draw [arrow] (pro4) -- (pro6);
			\draw [arrow] (pro5) -- (pro7);
			\draw [arrow] (pro6) -- (pro8);
			\draw [arrow] (pro7) -- (pro9);
			\draw [arrow] (pro8) -- (pro10);
			\draw [arrow] (pro9) -- (pro11);
			\draw [arrow] (pro10) -- (dec1);
			\draw [arrow] (pro11) -- (dec1);
			\draw [arrow] (dec1) -- node[anchor=east]  {yes}(io2);
			\draw [arrow] (dec1) -- node[anchor=west]  {no}(io3);
		\end{tikzpicture}
	\end{minipage}
	\caption{Entropic Causal Inference: Classical vs Quantum}
	\label{fig:classicVSquantum}
\end{figure*}
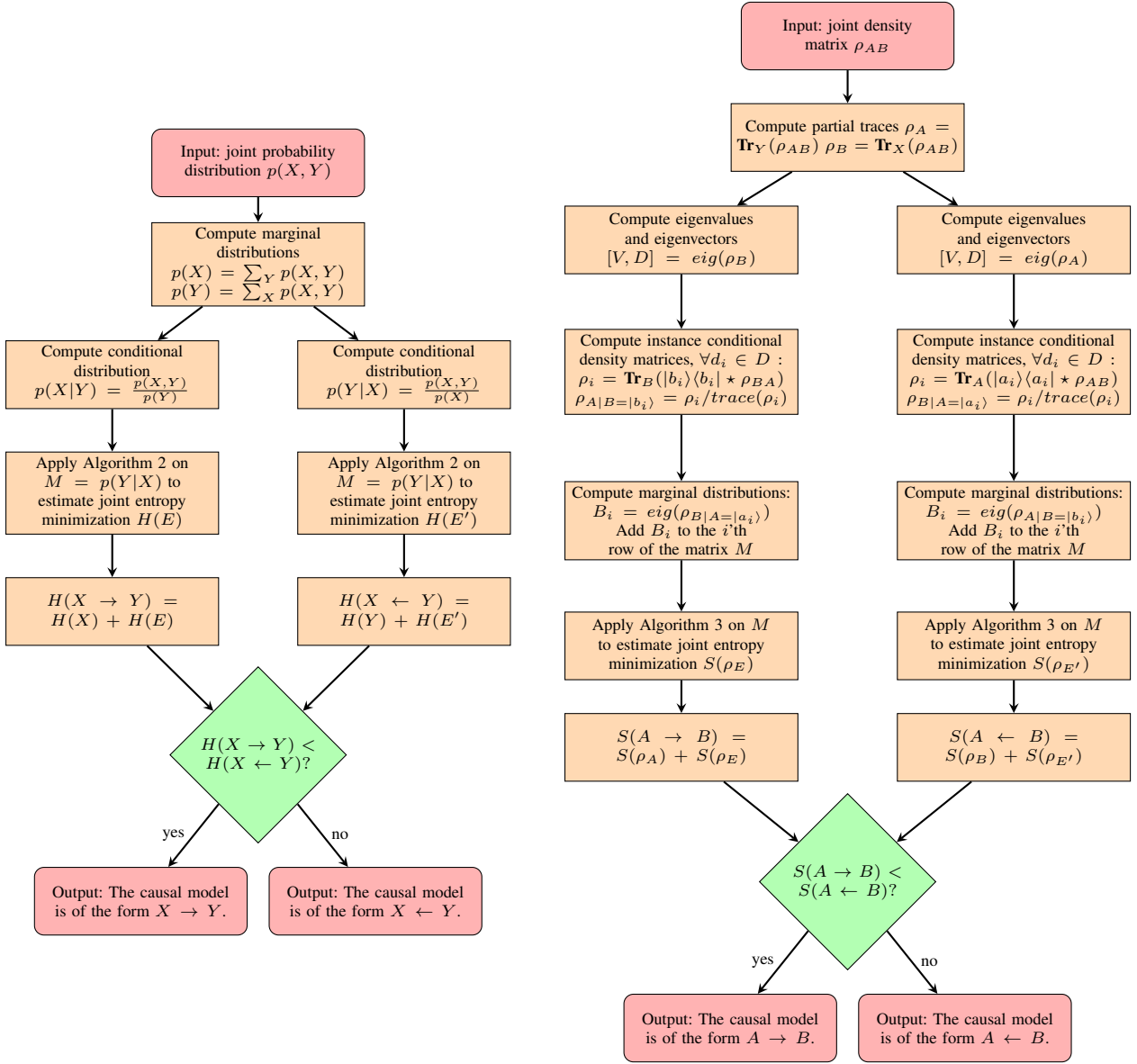

Using the diagonal approach, we obtain the corresponding joint density matrix $\rho_{AB}$.
\begin{equation*}
	\rho_{AB} = \left[
	\begin{array}{cccc}
		\frac{1}{16}  &  0 &   0 &   0\\
		0  &  \frac{3}{16} &   0 &   0\\
		0  &  0 &   \frac{5}{16} &   0\\
		0  &  0 &   0 &   \frac{7}{16}
	\end{array} \right]
\end{equation*}
This means that \qeci~captures both the classical and quantum entropic causal inference in a
unified framework. The parallels between classical and quantum entropic causal inference are illustrated in Figure \ref{fig:classicVSquantum}.
We now highlight similarities and differences between classical and quantum entropic causal inference.
 
\textbf{Key differences.} (1) Quantum entropic causal inference uses density matrices while the classical entropic causal inference uses probability distributions. (2) Quantum entropic causal inference uses partial traces while the classical entropic causal inference uses marginal probability distributions. (3) Quantum entropic causal inference uses instance conditional density matrices while the classical entropic causal inference uses conditional probability distributions. (4) Quantum entropic causal inference uses von Neumann entropy while the classical entropic causal inference uses Shannon entropy.

\textbf{Classical and quantum analogies.} (1) Both methods are built upon the assumption that the exogenous variable is simple (smaller) in the true causal direction (Assumption \ref{mainassump}).  (2) Both methods are built upon entropy as
a measure of simplicity for the causal discovery task. (3) Both methods are built upon  a greedy algorithm for searching minimum entropy of exogenous variables.

\section{Detailed workout for Quantum Symmetric Channel}\label{ex:qsca}

\begin{example}[Quantum Symmetric Channel]\label{ex:qsc}
	We consider a Quantum Symmetric Channel in the following model, in which the bit-flip error happens with certain probability. We consider the initial joint density as a superposition of $|0\rangle |0\rangle (|0\rangle |0\rangle)^\dagger$ with probability $q$, and $|1\rangle |1\rangle(|1\rangle |1\rangle)^\dagger$ with probability $1-q$. Further, the second qubit is transmitted over the quantum symmetric channel with the error probability $p$. After the transmission, the two qubits are labeled $A$ and $B$, respectively. 	The joint density operator $\rho_{AB}$ can be written as the superposition of the following density matrices, where $q$ is the probability that $A$ is in quantum state $|0\rangle$ and $1-q$ is the  probability that $A$ is in quantum state $|1\rangle$.
	
	
	\begin{equation*}
    \begin{aligned}
    \rho_{AB} = {}&
    q(1-p)*(|00\rangle \langle 00|)(|00\rangle \langle 00|)^\dagger\\
    &+qp*(|01\rangle \langle 01|)(|01\rangle \langle 01|)^\dagger\\
    &+(1-q)p*(|10\rangle \langle 10|)(|10\rangle \langle 10|)^\dagger\\
    &+(1-q)(1-p)*(|11\rangle \langle 11|)(|11\rangle \langle 11|)^\dagger.
    \end{aligned}
\end{equation*}
	
	In this case we already know that $A$ is the originator of the message. To verify this using our proposed method \qeci, assume that the joint density matrix $\rho_{AB}$ is given with $p=0.05$ and $q=0.4$. So, we have:
	\begin{equation*}
		\rho_{AB} = \left[
		\begin{array}{cccc}
			0.38 & 0 & 0 & 0\\
			0 & 0.02 & 0 & 0 \\
			0 & 0 & 0.03 & 0 \\
			0 & 0 & 0 & 0.57
		\end{array} \right]
	\end{equation*}
	Using \qeci~(Algorithm \ref{alg:qeci}) we have (here we trace the algorithm line by line  as follows):
	\begin{enumerate}
		\item Compute partial trace $\rho_A=\Tr_B(\rho_{AB})$: \begin{equation*}
			\rho_{A} = \left[
			\begin{array}{cc}
				0.4 & 0 \\
				0 & 0.6
			\end{array} \right]
		\end{equation*}
		\item Compute partial trace $\rho_B=\Tr_A(\rho_{AB})$: \begin{equation*}
			\rho_{B} = \left[
			\begin{array}{cc}
				0.41 & 0 \\
				0 & 0.59
			\end{array} \right]
		\end{equation*} 
		\item Compute $\rho_{BA}$ by reordering the entries of $\rho_{AB}$, as follows: \begin{equation*}
			\rho_{BA} = \left[
			\begin{array}{cccc}
				0.38 & 0 & 0 & 0\\
				0 & 0.03 & 0 & 0 \\
				0 & 0 & 0.02 & 0 \\
				0 & 0 & 0 & 0.57
			\end{array} \right]
		\end{equation*}%
		\item  Compute diagonal matrix $D$ of eigenvalues and matrix $V$ whose columns are the corresponding eigenvectors for $\rho_A$, so that $\rho_A*V = V*D$: $V = \begin{bmatrix}
			1 & 0 \\
			0 & 1
		\end{bmatrix}$, and 
		$D = \begin{bmatrix}
			0.4 & 0 \\
			0 & 0.6
		\end{bmatrix}$.
		\item \textbf{For} $d_0=0.4, d_1=0.6$ do:
		\item Compute pure states of eigenvectors $e_0=\begin{bmatrix}
			1 & 0\\
		\end{bmatrix}$ and $e_1=\begin{bmatrix}
			0 & 1\\
		\end{bmatrix}$ corresponding to the eigenvalues $d_0=0.4, d_1=0.6$, respectively, using the equation $x_i=e_i*e_i^T$, for $i=0,1$, where $e_i^T$ is the transpose of $e_i$: 
		
		$a_0=\begin{bmatrix}
			1  \\
			0\\
		\end{bmatrix}*\begin{bmatrix}
			1 & 0\\
		\end{bmatrix}=\begin{bmatrix}
			1 & 0 \\
			0 & 0\\
		\end{bmatrix}$
		
		$a_1=\begin{bmatrix}
			0  \\
			1\\
		\end{bmatrix}*\begin{bmatrix}
			0 & 1\\
		\end{bmatrix}=\begin{bmatrix}
			0 & 0 \\
			0 & 1\\
		\end{bmatrix}$
		\item Compute instance conditionals as defined in Definition \ref{def:cond_density}:
		
		$\rho_0=\begin{bmatrix}
			0.38 & 0 \\
			0 & 0.02\\
		\end{bmatrix}$, and $\rho_1=\begin{bmatrix}
			0.03 & 0 \\
			0 & 0.57\\
		\end{bmatrix}$
		\item Compute instance conditional densities (normalizing by the trace of $\rho_{B||a_i\rangle_A}$, for $i=0,1$): $\rho_{B||a_0\rangle_A}=\begin{bmatrix}
			0.95 & 0 \\
			0 & 0.05\\
		\end{bmatrix}$
		
		$\rho_{B||a_1\rangle_A}=\begin{bmatrix}
			0.05 & 0 \\
			0 & 0.95\\
		\end{bmatrix}$
		\item Compute marginal distributions 
		$B_i=eig(\rho_{B||a_i\rangle_A}$, for $i=0,1$:
		
		$B_0=\begin{bmatrix}
			0.05 & 0.95\\
		\end{bmatrix}$, and $B_1=\begin{bmatrix}
			0.05 & 0.95\\
		\end{bmatrix}$
		\item Add $B_i$'s to the i'th row of the matrix $M$:
		
		$M=\begin{bmatrix}
			0.05 & 0.95 \\
			0.05 & 0.95\\
		\end{bmatrix}$
		\item \textbf{End} of for loop.
		\item Apply Algorithm \ref{alg:jem} to estimate joint entropy minimization: $S(\rho_E)\approx 0.2864$
		\item Compute $S(A\to B)=S(\rho_A)+S(\rho_E)=0.2864+0.9710=1.2573$
		\item   Computes diagonal matrix $D$ of eigenvalues and matrix $V$ whose columns are the corresponding eigenvectors for $\rho_B$:
		
		$V = \begin{bmatrix}
			1 & 0 \\
			0 & 1
		\end{bmatrix}$, and 
		$D = \begin{bmatrix}
			0.41 & 0 \\
			0 & 0.59
		\end{bmatrix}$
		\item \textbf{For} $d_0=0.41, d_1=0.59$ do:
		\item Compute pure states of eigenvectors $e_0=\begin{bmatrix}
			1 & 0\\
		\end{bmatrix}$ and $e_1=\begin{bmatrix}
			0 & 1\\
		\end{bmatrix}$ corresponding to the eigenvalues $d_0=0.41, d_1=0.59$, respectively:
		
		$b_0=\begin{bmatrix}
			1  \\
			0\\
		\end{bmatrix}*\begin{bmatrix}
			1 & 0\\
		\end{bmatrix}=\begin{bmatrix}
			1 & 0 \\
			0 & 0\\
		\end{bmatrix}$
		
		$b_1=\begin{bmatrix}
			0  \\
			1\\
		\end{bmatrix}*\begin{bmatrix}
			0 & 1\\
		\end{bmatrix}=\begin{bmatrix}
			0 & 0 \\
			0 & 1\\
		\end{bmatrix}$
		\item Compute instance conditionals as defined in Definition \ref{def:cond_density}:
		
		$\rho_0=\begin{bmatrix}
			0.38 & 0 \\
			0 & 0.03\\
		\end{bmatrix}$, and $\rho_1=\begin{bmatrix}
			0.02 & 0 \\
			0 & 0.57\\
		\end{bmatrix}$
		\item Compute instance conditional density matrices (normalizing by the trace of $\rho_{A||b_i\rangle_B}$, for $i=0,1$):
		
		$\rho_{A||b_0\rangle_B}=\begin{bmatrix}
			0.9268 & 0 \\
			0 & 0.0732\\
		\end{bmatrix}$
		
		$\rho_{A||b_1\rangle_B}=\begin{bmatrix}
			0.0339 & 0 \\
			0 & 0.9661\\
		\end{bmatrix}$
		\item Compute marginal distributions 
		$B_i=eig(\rho_{A||b_i\rangle_B}$, for $i=0,1$:
		
		$B_0=\begin{bmatrix}
			0.0732 & 0.9268\\
		\end{bmatrix}$
		$B_1=\begin{bmatrix}
			0.0339 & 0.9661\\
		\end{bmatrix}$
		\item Add $B_i$'s to the i'th row of the matrix $M$:
		
		$M=\begin{bmatrix}
			0.0732 & 0.9268  \\
			0.0339 & 0.9661\\
		\end{bmatrix}$
		\item \textbf{End} of for loop.
		\item Apply Algorithm \ref{alg:jem} to estimate joint entropy minimization: $S(\rho_{E'})\approx 0.4505$
		\item Compute $S(A\gets B)=S(\rho_B)+S(\rho_{E'})=0.4505+0.9765=1.4270$
		\item Compare $S(A\to B)$ and $S(A\gets B)$: $S(A\to B)<S(A\gets B)$
		\item \textbf{return}: The causal model is of the form $A\to B.$
	\end{enumerate}
\end{example}





\end{document}